\documentclass[showpacs,preprintnumbers,amsmath,amssymb]{revtex4}

\usepackage{graphicx}
\usepackage{dcolumn}
\usepackage{bm}
\usepackage{epsfig}

\newcommand{\dhd}{{\textstyle d}
\lower.03ex\hbox{\kern-0.40em$^{\scriptstyle-}$}\kern-0.08em{}}  

\newcommand{\half}{{1\over 2}}
\newcommand{\bu}{{\bullet}}

\begin{document}

\preprint{JLAB-THY-06-541}

\title{
Quark contribution to the small-$x$ evolution of color
dipole}

\author{Ian Balitsky}
\affiliation{
Physics Dept., ODU, Norfolk VA 23529, \\
and \\
Theory Group, Jlab, 12000 Jeffeson Ave, Newport News, VA 23606
}
\email{balitsky@jlab.org}

\date{\today}

\begin{abstract}
The small-$x$ deep inelastic scattering in the saturation region is governed by 
the non-linear evolution of Wilson-lines operators. 
In the leading logarithmic approximation it is given by the BK
equation for the evolution of color dipoles. In the NLO the nonlinear equation gets contributions from quark and 
gluon loops. In this paper I calculate the quark-loop contribution to small-x 
evolution of Wilson lines in the NLO. 
It turns out that there are no new operators at the one-loop level - just as at the tree level, the high-energy scattering can be described in terms 
of Wilson lines. In addition, from the analysis of quark loops 
I find that the argument of coupling constant in the BK equation is determined by the size of the parent dipole
rather than by the size of produced dipoles. These results are to be supported by future calculation of gluon loops.
\end{abstract}


\maketitle

\section{\label{sec:in}Introduction }
At high energies the particles move very fast along straight lines, hence they can be described by Wilson lines $U^\eta(x_\perp)$ - gauge factors ordered along straight-line classical trajectory of the particle moving with rapidity $\eta$ at the 
transverse impact
parameter $x_\perp$ (for a review, see \onlinecite{mobzor}). For deep inelastic scattering, the propagation of a quark-antiquark pair moving along straight lines and separated by a distance in the transverse direction can be approximated by the color dipole $U(x_\perp)U^\dagger(y_\perp)$ - two Wilson lines ordered along the direction 
collinear to quarks' velocity. The structure function of a hadron is then
proportional to a matrix element of the color dipole operator 
\begin{equation}
{\cal U}^\eta(x_\perp,y_\perp)=1-{1\over N_c}
{\rm Tr}\{U^\eta(x_\perp)U^{\dagger\eta}(y_\perp)\}
\label{fla1}
\end{equation}
switched between the target's states ($N_c=3$ for QCD).  Approximately,  the gluon parton density is
\begin{equation}
x_BG(x_B,\mu^2=Q^2)~
\simeq ~\left.\langle p|~{\cal U}^\eta(x_\perp,0)|p\rangle
\right|_{x_\perp^2=Q^{-2}}
\label{fla2}
\end{equation}
where $\eta=\ln{1\over x_B}$ and $x_B={Q^2\over 2(p\cdot q)}$ is the Bjorken variable.

The small-x behavior of the structure functions is governed by the 
small-$x$ evolution of color dipoles\cite{mu94,nnn}. For sufficiently small dipoles 
$x_\perp^2\sim Q^{-2}$ so $\alpha_s(Q)\ll $1 and we can use pQCD. At high 
(but not asymptotic) energies we can use the leading logarithmic approximation (LLA)
where  $ \alpha_s\ll 1,~ \alpha_s\ln x_B\sim 1$. 
In the LLA, the high-energy amplitudes in pQCD
are described by the BFKL equation\cite{bfkl} leading to the power behavior 
$F_2(x_B)\sim x_B^{-12{\alpha_s\over\pi}\ln 2}$. However, the example of DIS from very large nuclei shows that the BFKL equation is not sufficient to describe the small-$x$ behavoir of structure functions even in the LLA. Indeed, at sufficiently large atomic number $A$ we get an additional parameter  $\alpha_sA^{1/6}\sim 1$ which must be taken into account exactly to all orders of the expansion in this parameter. The situation is essentially semiclassical: we have $\alpha_s\ll 1$ and $\alpha_sF_{\mu\nu}\sim 1$ where $F_{\mu\nu}$ is the strong field of the nucleus gluon cloud. Thus we need the LLA in the semiclassical QCD
(sQCD): $ \alpha_s\ll 1,~ \alpha_s\ln x_B\sim 1,~  \alpha_sF_{\mu\nu}\sim 1$. 
This situation appears to be general for sufficiently low $x_B$: even for the proton,
 where we do not have the large parameter $A$ to start with, the power 
behavior of gluon parton density will lead to the huge number of partons in the target  
leading to the state of saturation\cite{saturation} described by 
Color Glass Condensate in sQCD\cite{lvmodel,jimwalk}.

The LLA  evolution equation for the color dipoles is non-linear\cite{npb96,yura}:
\begin{eqnarray}
&&\hspace{-1mm}
{d\over d\eta}~{\cal U}(x,y)~=~
{\alpha_sN_c\over 2\pi^2}\!\int\!d^2z~ {(x-y)^2\over(x-z)^2(z-y)^2}
[{\cal U}(x,z)+{\cal U}(y,z)-{\cal U}(x,y)-{\cal U}(x,z){\cal U}(z,y)]
\label{bk}
\end{eqnarray}
The first three terms correspond to the linear BFKL evolution and describe the parton emission while the last term is responsible for the parton annihilation. For sufficiently high $x_B$ the parton emission balances the parton annihilation so the partons reach the state of saturation with
the characteristic transverse momentum $Q_s$ growing with $x_B$ as $e^{c\ln x_B}$.
The argument of the coupling constant in Eq. (\ref{bk}) is left undetermined in the LLA, and usually it is set by hand to be $Q_s$. Careful analysis of this argument is very important  from both theoretical and experimental points of view. From the theoretical viewpoint, we need to know whether the
coupling constant is determined by the size of the original dipole $|x-y|$ or of the size of the produced dipoles $|x-z|$ and/or $|z-y|$ since we may get a very different behavior
of the solutions of the equation (\ref{bk}) (although first numerical simulations indicate a slow dependence of the cross section on the choice of the scale\cite{scaledep}). On the experimental side, the cross section is  proportional to some power of the coupling constant so  the argument determines 
how big (or how small) is the cross section. The typical argument of $\alpha_s$ is 
the characteristic transverse momenta of the process. For high enough energies, they are believed to be of order of the saturation scale $Q_s$ which is $\sim 2\div 3$ GeV for the LHC collider. Thus, we see that even the difference between $\alpha(Q_s)$ and 
$\alpha(2Q_s)$ can make a huge impact on the cross section.

The argument of the coupling constant cannot be determined in the LLA 
so the next-to-leading order (NLO) calculation is in order.
In the next-to-leading order the non-linear equation (\ref{bk}) looks as follows ($x,y,z..$ are the transverse coordinates)
\begin{eqnarray}
&&\hspace{-1mm}
{d\over d\eta}~{\rm Tr}\{U_x U^{\dagger}_y\}~=~
{1\over 2\pi^2}\!\int\!d^2z\Big(\alpha_s {(x-y)^2\over(x-z)^2(z-y)^2}
+~\alpha_s^2 K_{\rm NLO}(x,y,z)\Big)[
{\rm Tr}\{U_xU^\dagger_z\}{\rm Tr}\{U_zU^\dagger_y\}-
N_c{\rm Tr}\{U_zU^\dagger_y\}]
\nonumber\\
&&\hspace{-1mm}+~
\alpha_s^2\!\int\!d^2z d^2z' \Big(K_4(x,y,z,z')\{U_x,U^\dagger_{z'},U_z,U^\dagger_y\}
+~K_6(x,y,z,z')\{U_x,U^\dagger_{z'},U_{z'},U_z,U^\dagger_z,U^\dagger_y\}\Big)
\label{proekt}
\end{eqnarray}
where $K_{\rm NLO}$ is the next-to-leading order correction to the dipole kernel and 
$K_4$ and $K_6$ are the coefficients in front of the (tree) four- and six-Wilson line operators with arbitrary white arrangements of color indices. 
Note that $K_{\rm NLO}$ must describe the non-forward NLO BFKL contribution
found recently in Ref. \onlinecite{nfnlobfkl}.
(The contribution  $\sim K_6$ proportional to six Wilson-line operators  the was obtained in Ref. \onlinecite{balbel}).
The calculation of the 
quark part of the kernel is performed in the present paper and the 
last remaining part of Eq. (\ref{proekt}) - the calculation of the gluon part of $K_{\rm NLO}$ and $K_4$ - is in progress. 

It should be mentioned that  NLO result does not lead automatically to
the argument of coupling constant in front of the leading term in Eq. \ref{proekt}. 
 In order to get this argument, we can use the 
 renormalon-based approach\cite{renormalons}: first we get  the quark part 
 of the running coupling constant  coming from the bubble chain of quark loops and then  make a conjecture that the gluon part 
 of the $\beta$-function will follow that pattern (see the discussion
 in Refs \onlinecite{blm,braunbeneke94}).

  As we demonstrate below, the result is that the value of coupling constant is determined by the size of the original dipole rather than the size of the produced dipoles:
\begin{eqnarray}
&&\hspace{-1mm}
{d\over d\eta}~{\cal U}(x,y)~=~
{\alpha_s((x-y)^2)N_c\over 2\pi^2}\!\int\!d^2z~ {(x-y)^2\over(x-z)^2(z-y)^2}
[{\cal U}(x,z)+{\cal U}(y,z)-{\cal U}(x,y)-{\cal U}(x,z){\cal U}(z,y)]+...
\label{bkarg}
\end{eqnarray}
The paper is organized as follows. In  Sect. 2 I recall the derivation of the 
BK equation in the leading order in $\alpha_s$. In Sect. 3, which is central to the paper,
I calculate the quark contribution to the small-$x$ evolution kernel of Wilson-line operators. 
In Sect. 4  I present the arguments 
that the coupling constant in the BK equation is determined by the size $(x-y)_\perp$ of the parent dipole.
The light-cone expansion of the quark-loop propagator is performed in the Appendix.

\section{Derivation of the BK equation}
Before discussion of the small-x evolution of color dipole in the next-to-leading approximation it is instructive to recall the derivation of the leading-order (BK)
evolution equation. 
As discussed in the Introduction, the dependence of the structure functions 
on $x_B$ comes from the dependence of Wilson-line operators   
\begin{equation}
U^\eta(x_\perp)={\rm Pexp}\Big\{ig\int_{-\infty}^\infty\!\!  du ~p^\eta_\mu ~A^\mu(up^\eta+x_\perp)\Big\},~~~~
p^\eta\equiv p_1+e^{-\eta}p_2
\label{defu}
\end{equation}
on the slope of the supporting line. Here $p_1$ and $p_2$ are the light-like
vectors such that $q=p_1-x_B p_2$ and  $p=p_2+{m^2\over s}p_1$ where
$p$ is the momentum of the target and $m$ is the mass. Throughout the paper, we use the 
Sudakov variables $p=\alpha p_1+\beta p_2 +p_\perp$ and the notations 
$x_\bullet\equiv x_\mu p_1^\mu$ and $x_\ast\equiv x_\mu p_2^\mu$ related to
the light-cone coordinates: $x_\ast=x^+\sqrt{s/2},~x_\bullet=x^-\sqrt{s/2}$. 

\begin{figure}
\includegraphics[width=0.98\textwidth]{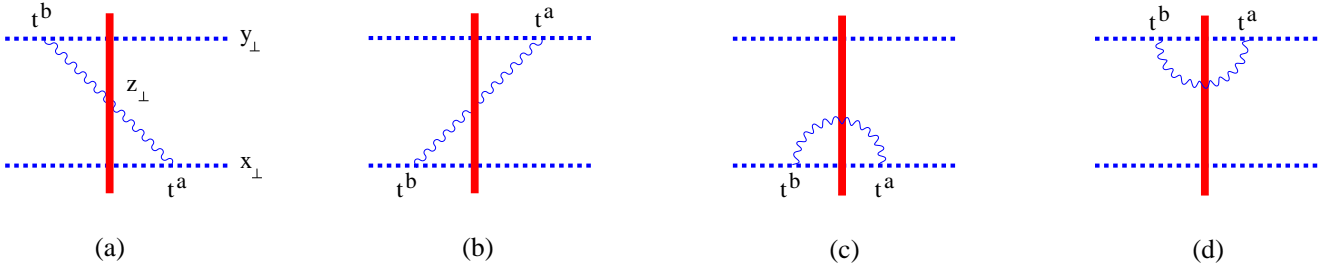}
\caption{Leading-order diagrams for the small-$x$ evolution of color dipole\label{bkevol}.}
\end{figure}

To find the evolution of the color dipole (\ref{fla1}) with respect to the slope of the 
Wilson lines in the leading log approximation, we consider the matrix element of the color dipole between (arbitrary) target states and integrate over the gluons with rapidities $\eta_1>\eta>\eta_2=\eta_1-\Delta\eta$ leaving the gluons with $\eta<\eta_2$ as
the background field (to be integrated over later).
In the frame of gluons with $\eta\sim\eta_1$ the fields with
$\eta<\eta_2$ shrink to a pancake and we obtain the four diagrams shown in Fig. 
\ref{bkevol}

The (background-Feynman) gluon propagator in a shock-wave external field 
has the form\cite{npb96,prd99}
\begin{eqnarray}
&&\hspace{-2mm}\langle A^a_{\mu}(x)A^b_{\nu}(y)\rangle ~=~
\theta(x_\ast y_\ast)
(x|{g_{\mu\nu}\delta^{ab}\over i(p^2+i\epsilon)}|y)~
\label{gluprop}\\
&&\hspace{-2mm}
-~\theta(x_\ast)\theta(-y_\ast)
\!\int_0^\infty\!\dhd\alpha~ {e^{-i\alpha(x-y)_\bullet}\over 4\alpha^2}
(x_\perp|e^{-i{p_\perp^2\over\alpha s}x_\ast}
\Big[2\alpha g_{\mu\nu}U+{4\over s}
(i\partial_{\mu}p_{2\nu}U-p_{2\mu}i\partial_{\nu}U)-
{4p_{2\mu}p_{2\nu}\over\alpha s^2}\partial_{\perp}^2U\Big]
e^{i{p_\perp^2\over\alpha s}y_\ast}
|y_\perp)^{ab} 
\nonumber \\
&&\hspace{-2mm}-~\theta(-x_\ast)\theta(y_\ast)
\!\int_0^\infty\!\dhd\alpha~ {e^{i\alpha(x-y)_\bullet}\over 4\alpha^2}
(x_\perp|e^{-i{p_\perp^2\over\alpha s}x_\ast}
\Big[2\alpha g_{\mu\nu}U^{\dagger}-{4\over s}
(i\partial_{\mu}U^{\dagger}p_{2\nu}-
p_{2\mu}i\partial_{\nu}U^{\dagger})-
{4p_{2\mu}p_{2\nu}\over\alpha s^2}\partial_{\perp}^2U^{\dagger}
\Big]
e^{i{p_\perp^2\over\alpha s}y_\ast}
|y_\perp)^{ab} \nonumber
\end{eqnarray}
where $\partial_{\perp}^2\equiv -\partial_i\partial^i$. 
Hereafter use Schwinger's notations 
$(x|F(p)|y)\equiv \int\!\dhd p~F(p)~e^{-ip\cdot(x-y)}$ and 
$(x_\perp|F(p_\perp)|y)\equiv \int\!\dhd p~e^{i(p,x-y)_\perp}$ ( the scalar product of the four-dimensional vectors in our notations is $x\cdot y={2\over s}(x_\ast y_\bullet+x_\ast y_\bullet)-(x,y)_\perp$). We obtain
\begin{eqnarray}
&&\hspace{-26mm}\Big\{\!\int_0^\infty \! du \int^0_{-\infty} \! dv~A^a_\bu(up^{\eta_1}+x_\perp)
A^b_\bu(vp^{\eta_1}+y_\perp)\Big\}_{\rm Fig. \ref{bkevol}a}
\nonumber\\
&&\hspace{-26mm}=~-2\alpha_s
\int_{e^{-\eta_2}}^\infty\!{d\alpha\over\alpha}
(x_\perp|{1\over p_\perp^2+\alpha^2e^{-2\eta_1}s}\partial_\perp^2 
U^{ab}{1\over p_\perp^2+\alpha^2 e^{-2\eta_1}s}|y_\perp)
\label{bk1}
\end{eqnarray}
Formally, the integral over $\alpha$ diverges at the lower limit, but since we integrate over the rapidities $\eta>\eta_2$ we get in the LLA 
\begin{eqnarray}
&&\hspace{-26mm}\Big\{\!\int_0^\infty \! du \int^0_{-\infty} \! dv~A^a_\bu(up_\eta+x_\perp)
A^b_\bu(vp_\eta+y_\perp)\Big\}_{\rm Fig. \ref{bkevol}a}
\nonumber\\
&&\hspace{-26mm}=~-2\alpha_s\Delta\eta
(x_\perp|{1\over p_\perp^2}\partial_\perp^2 
U^{ab}{1\over p_\perp^2}|y_\perp)~=~
-2\alpha_s\Delta\eta\!\int\! d^2z_\perp (x_\perp|{p_i\over p_\perp^2}|z_\perp)
(2U_z-U_x-U_y)^{ab}(z_\perp|{p_i\over p_\perp^2}|y_\perp)
\label{bk2}
\end{eqnarray}
and therefore
\begin{eqnarray}
&&\hspace{-2mm}\Big\{U_x\otimes U^\dagger_y\Big\}_{\rm Fig. \ref{bkevol}a}^{\eta_1}
\nonumber\\
&&\hspace{-2mm}=~-{\alpha_s\over 2\pi^2}\Delta\eta~ \{t^aU_x\otimes t^bU^\dagger_y\}^{\eta_2}
\!\int\! d^2z_\perp {(x-z,y-z)_\perp\over (x-z)_\perp^2(y-z)_\perp^2}
(2U_z^{\eta_2}-U_x^{\eta_2}-U_y^{\eta_2})^{ab}
\label{bk3}
\end{eqnarray}
The contribution of the diagram in Fig.  \ref{bkevol}b is obtained from Eq. (\ref{bk3})
by the replacement $t^aU_x\otimes t^bU^\dagger_y \rightarrow U_x t^b\otimes U^\dagger_y t^a$, $x\leftrightarrow y$ and the two remaining diagrams are obtained from
Eq. \ref{bk2} by taking $y=x$ (Fig. \ref{bkevol}c) and  $x=y$ (Fig. \ref{bkevol}d).
Finally, one obtains
\begin{eqnarray}
&&\hspace{-2mm}\Big\{U_x\otimes U^\dagger_y\Big\}_{\rm Fig. \ref{bkevol}}^{\eta_1}
~=~-{\alpha_s\Delta\eta\over 2\pi^2} \{t^aU_x\otimes t^bU^\dagger_y
+ U_x t^b\otimes U^\dagger_y t^a\}^{\eta_2}
\!\int\! d^2z_\perp {(x-z,y-z)_\perp\over (x-z)_\perp^2(y-z)_\perp^2}
(2U_z^{\eta_2}-U_x^{\eta_2}-U_y^{\eta_2})^{ab}
\nonumber\\
&&\hspace{-2mm}+~{\alpha_s\Delta\eta\over \pi^2}
 \{t^aU_xt^b\otimes U^\dagger_y\}^{\eta_2}
\!\int\! {d^2z_\perp \over (x-z)_\perp^2}
(U_z^{\eta_2}-U_x^{\eta_2})^{ab}
+{\alpha_s\Delta\eta\over \pi^2} \{U_x\otimes t^bU^\dagger_yt^a\}^{\eta_2}
\!\int\! {d^2z_\perp\over (y-z)_\perp^2}
(U_z^{\eta_2}-U_y^{\eta_2})^{ab}
\label{bk4}
\end{eqnarray}
For the color dipole (\ref{fla1}) one easily gets the BK equation (\ref{bk}).
\section{Quark contribution to the NLO BK kernel}
\subsection{Quark loop in the momentum representation}

\begin{figure}
\includegraphics[width=0.9\textwidth]{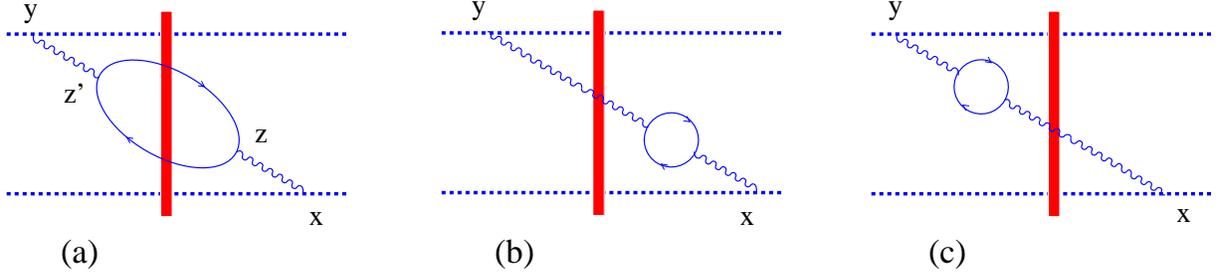}
\caption{Quark-loop contribution to the gluon propagator in a shock-wave background \label{kvloop1}.}
\end{figure}

There are two types of quark contribution in the NLO: with quarks in the loop 
interacting with the shock wave (see  Fig. \ref{kvloop1}a) or without  
 ( Fig. \ref{kvloop1}b). (In principle, there could have been the contribution coming from the quark loop which lies entirely in the shock wave, but we will demonstrate below that
it vanishes).
 
The quark propagator in a shock-wave background has the form \cite{npb96}:
\begin{eqnarray}
&&\hspace{-16mm}
\psi(x)\bar{\psi}(y)~=~
\theta(x_\ast y_\ast)(x|{i\not\! p\over p^2+i\epsilon}|y)
\nonumber\\ 
&&\hspace{-16mm}+~\theta(x_\ast)\theta(-y_\ast)
\!\int_0^\infty\!{\dhd\alpha\over 2\alpha^2 s}~e^{-i\alpha(x-y)_\bullet}~
(x_\perp|(\alpha\hat{p}_1+\hat{p}_\perp)
e^{-i{p_\perp^2\over\alpha s}x_\ast}
\hat{p}_2Ue^{i{p_\perp^2\over\alpha s}y_\ast}(\alpha\hat{p}_1+\hat{p}_\perp)|y_\perp)
\nonumber\\ 
&&\hspace{-16mm}
-~\theta(-x_\ast)\theta(y_\ast)\!\int_{-\infty}^0{\dhd\alpha\over 2\alpha^2 s}~e^{-i\alpha(x-y)_\bullet}~
(x_\perp|(\alpha\hat{p}_1+\hat{p}_\perp)
e^{-i{p_\perp^2\over\alpha s}x_\ast}
\hat{p}_2U^\dagger e^{i{p_\perp^2\over\alpha s}y_\ast}(\alpha\hat{p}_1+\hat{p}_\perp)|y_\perp)
\label{kvpropagator}
\end{eqnarray}

Multiplying two propagators one gets at $x_\ast>0, y_\ast<0$
\begin{eqnarray}
&&\hspace{-16mm}
\bar{\psi}t^a\!\not\! p_1\psi(x)\bar{\psi}(y)\!\not\! p_1\psi(y)~=~
\nonumber\\ 
&&\hspace{-16mm}=~
\!\int_0^\infty\!{\dhd\alpha\over 16\pi\alpha^3 }\!\int_0^1\! {dv\over \bar{v}^2 v^2}
~e^{-i\alpha(x-y)_\bullet}~{\rm tr}{\rm Tr}
(x_\perp|\not\!p_\perp
e^{-i{p_\perp^2\over\alpha v s}x_\ast}
Ue^{i{p_\perp^2\over\alpha vs}y_\ast}\!\not\! p_\perp|y_\perp)
~t^b(y_\perp|\!\not\! p_\perp
e^{i{p_\perp^2\over\alpha \bar{v}s}y_\ast}
U^\dagger e^{i{p_\perp^2\over\alpha \bar{v}s}y_\ast}\!\not\! p_\perp|x_\perp)
\label{kvlup}
\end{eqnarray}
where tr stands for the trace over spinor indices.
Therefore the quark-loop contribution to the gluon propagator is
\begin{eqnarray}
&&\hspace{-2mm}\langle A^a_\bu(x)
A^b_\bu(y)\rangle
\nonumber\\
&&\hspace{-2mm}=~2\alpha^2_sn_f
\int_0^\infty\!{d\alpha\over\alpha^3}\!\int\! d^2z d^2z'\!
\int \dhd^2k_1\dhd^2k'_1\dhd^2k_2\dhd^2k'_2~e^{i(k_1,x-z)_\perp+i(k'_1,x-z')_\perp
-i(k_2,y-z)_\perp-i(k'_2,y-z')_\perp}
\nonumber\\
&&\hspace{-2mm}
{\rm Tr}\{t^aU_zt^bU^\dagger_{z'}\}
\int_0^1\! dv~
{(k_1,k_2)(k'_1,k'_2)
+(k_1,k'_1)(k_2,k'_2)-(k_1,k'_2)(k'_1,k_2)
\over [(k_1+k'_1)^2\bar{v}v-k^2_1\bar{v}-{k'_1}^2v][(k_2+k'_2)^2\bar{v}v-
k^2_2\bar{v}-{k'_2}^2v]}  
 \nonumber\\
&&\hspace{-2mm}
\times~
 \Big[e^{-i{(k_1+k'_1)_\perp^2\over\alpha s}x_\ast}
 -e^{-i\Big({k_1^2\over v}+{{k'_1}^2\over\bar v}\Big){x_\ast\over\alpha s}}\Big]
\Big[e^{i{(k_2+k'_2)_\perp^2\over\alpha s}y_\ast}
 -e^{i({k_2^2\over v}+{{k'_2}^2\over\bar v}){y_\ast\over\alpha s}}\Big]
\label{kvlup2}
\end{eqnarray}
and one obtains
 the contribution of the diagram in Fig. \ref{kvloop1}a in the form 

\begin{eqnarray}
&&\hspace{-26mm}\Big\{\!\int_0^\infty \! du \int^0_{-\infty} \! dv~A^a_\bu(up_1+x_\perp)
A^b_\bu(vp_1+y_\perp)\Big\}_{\rm Fig. \ref{kvloop1}a}
\nonumber\\
&&\hspace{-26mm}=~-8\alpha^2_sn_f
\int_0^\infty\!{d\alpha\over\alpha}\!\int\! dz dz'\!
\int \dhd^2k_1\dhd^2k'_1\dhd^2k_2\dhd^2k'_2~e^{i(k_1,x-z)_\perp+i(k'_1,x-z')_\perp
-i(k_2,y-z)_\perp-i(k'_2,y-z')_\perp}
\nonumber\\
&&\hspace{-1mm}
{\rm Tr}\{t^aU_zt^bU^\dagger_{z'}\}
\int_0^1\! dv~
{(k_1,k_2)(k'_1,k'_2)
+(k_1,k'_1)(k_2,k'_2)-(k_1,k'_2)(k'_1,k_2)
\over (k_1+k'_1)^2(k_2+k'_2)^2
(k^2_1v+{k'_1}^2\bar{v})(k^2_2v+{k'_2}^2\bar{v})}   
\label{kvlup1a}
\end{eqnarray}
where $n_f$ is a number of light quarks 
 ($n_f=3$ for the momenta $Q_s\sim 1\div 2$ GeV)
 and Tr stands for the trace over color indices. The variable $v$ is the fraction of the gluon's momentum $\alpha$ carried by the quark.

To calculate this diagram we use the dimensional regularization and
change the dimension of the transverse space to $d=2-2\epsilon$. 
The calculation yields
\begin{eqnarray}
&&\hspace{-6mm}
{\rm Tr}\{U_x U^\dagger_y\}_{\rm Fig. \ref{kvloop1}a}
\nonumber\\
&&\hspace{-6mm}
=~-{4\alpha^2_s\over\pi}n_f
\Delta\eta
{\rm Tr}\{t^aU_xt^bU^{\dagger}_y\}
\int\! \dhd^dp~\dhd^dq~\dhd^dq' ~
{e^{i(p,x)_\perp-i(p-q-q',y)_\perp}\over p^2(p-q-q')^2}
\!\int_0^1 \! dv du~
\Bigg[-
(q+q')^2{\bar{u}u\Gamma(\epsilon) \mu^{2\epsilon}\over 
(P^2\bar{v} v+Q^2 \bar{u} u)^\epsilon}
\nonumber\\
&&\hspace{-6mm}
+~{\Gamma(1+\epsilon) \mu^{2\epsilon}\over 
(P^2\bar{v} v+Q^2 \bar{u} u)^{1+\epsilon}}
\Big\{P^2[\bar{v}v(q,q')-\bar{u}uQ^2+2\bar{u}u\bar{v}v(q^2+{q'}^2)]
-2\bar{u}u\bar{v}v(P,q)(P,q')
\nonumber\\
&&\hspace{-6mm}
+~\bar{u}u(1-2u)[\bar{v}v(q,q')(P,q+q')
+\bar{v}q^2(P,q')+v{q'}^2(P,q)]
+\bar{u}^2u^2Q^2(q+q')^2\Big\}\Bigg]
{\rm Tr}\{t^a
U(q)t^b U^{\dagger}(q')\}
\label{vklad1}
\end{eqnarray}

where $P\equiv p-(q+q')u$,
 $Q^2=q_\perp^2\bar{v}+{q'}_\perp^2 v$ and we use the notation
 $\int\!\dhd^d p_\perp\equiv \!\int\! {dp_\perp\over(2\pi)^d}$.

\begin{figure}
\includegraphics[width=0.9\textwidth]{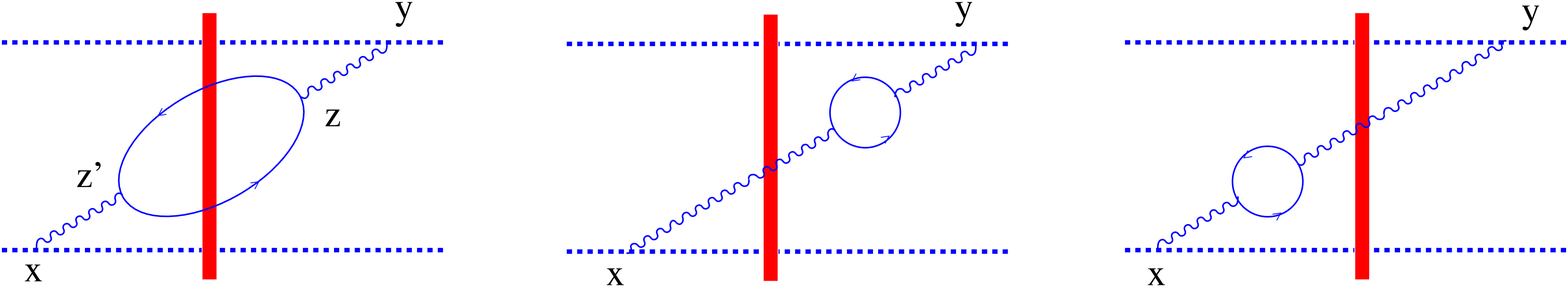}
\caption{\label{kvloop2}.}
\end{figure}

\begin{figure}
\includegraphics[width=0.9\textwidth]{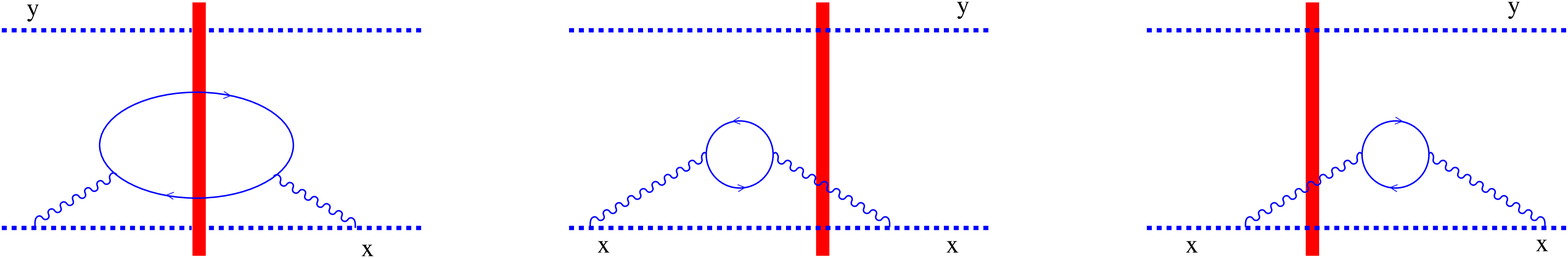}
\caption{ \label{kvloop3}.}
\end{figure}

\begin{figure}
\includegraphics[width=0.9\textwidth]{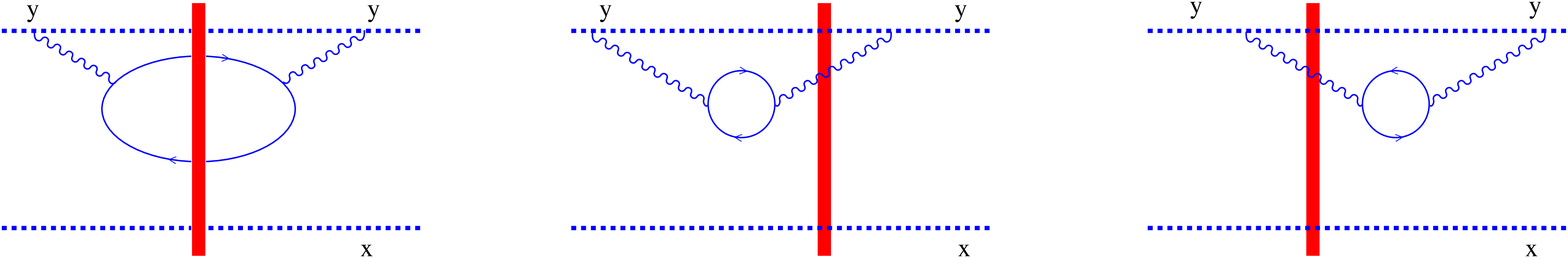}
\caption{ \label{kvloop4}.}
\end{figure}

The contribution of two diagrams in Fig. \ref{kvloop1}b,c is
\begin{eqnarray}
&&\hspace{-16mm}
2n_f\alpha^2_s\Delta\eta
{\rm tr}\{t^aU_x t^bU^{\dagger}_y\}{\mu^{2\epsilon}\over \pi} B(2-\epsilon,2-\epsilon)
(x_\perp|{1\over p_\perp^2}\Big\{
{\Gamma(\epsilon)\over (p_\perp^2)^\epsilon},\partial_\perp^2U^{ab}\Big\}
{1\over p_\perp^2}|y_\perp)
\label{kvloopbc}
\end{eqnarray}
where $B(a,b)=\Gamma(a)\Gamma(b)/\Gamma(a+b)$.
The contribution of diagrams in Fig. \ref{kvloop2} is obtained from
the sum of Eq. (\ref{vklad1}) and (\ref{kvloopbc})  by the
 $x\leftrightarrow y$ replacement in the coefficient in front of 
 ${\rm tr}\{t^aU_x t^bU^{\dagger}_y\}$
and the contribution of the diagram in Fig. \ref{kvloop3}  by taking $y\rightarrow x$ in this coefficient and changing the sign.
Similarly, the diagram in Fig. \ref{kvloop4} is obtained by taking $x\rightarrow y$.
The sum of all diagrams has the form
\begin{eqnarray}
&&\hspace{-3mm}
{\rm Tr}\{U_x U^\dagger_y\}~=
-{\alpha^2_s\over \pi}n_f
\Delta\eta
{\rm Tr}\{t^aU_x t^bU^{\dagger}_y\}
\label{vesvklad1}
\\
&&\hspace{-3mm}
\times~\Bigg(
\int\! {\dhd^dp~\dhd^dq~\dhd^dq' \over p^2(p-q-q')^2} ~
(e^{i(p,x)_\perp}-e^{i(p,y)_\perp})
(e^{-i(p-q-q',x)_\perp}-e^{-i(p-q-q',y)_\perp})
\!\int_0^1 \! dudv ~
\Big[
(q+q')^2{4\bar{u} u\Gamma(\epsilon) \mu^{2\epsilon}\over 
(P^2\bar{v} v+Q^2 \bar{u} u)^\epsilon}
\nonumber\\
&&\hspace{-3mm}
-~4{\Gamma(1+\epsilon) \mu^{2-d}\over 
(P^2\bar{v} v+Q^2 \bar{u} u)^{1+\epsilon}}
\Big\{P^2[\bar{v}v(q,q')-\bar{u}uQ^2+2\bar{u}u\bar{v}v(q^2+{q'}^2)]
-2\bar{u}u\bar{v}v(P,q)(P,q')
\nonumber\\
&&\hspace{-3mm}
+~\bar{u}u(1-2u)[\bar{v}v(q,q')(P,q+q')+\bar{v}q^2(P,q')+v{q'}^2(P,q)]
+\bar{u}^2u^2Q^2(q+q')^2\Big\}\Big]
{\rm Tr}\{t^aU(q_\perp)t^bU^\dagger(q'_\perp)\}
\nonumber\\
&&\hspace{-3mm}
-~
2\mu^{2\epsilon} B(2-\epsilon,2-\epsilon)
\Big[2(x_\perp|{1\over p_\perp^2}\Big\{
{\Gamma(\epsilon)\over p_\perp^{2\epsilon}},\partial_\perp^2U^{ab}\Big\}
{1\over p_\perp^2}|y_\perp)-(x_\perp|{1\over p_\perp^2}\Big\{
{\Gamma(\epsilon)\over p_\perp^{2\epsilon}},\partial_\perp^2U^{ab}\Big\}
{1\over p_\perp^2}|x_\perp)
\nonumber\\
&&\hspace{-3mm}
-~(y_\perp|{1\over p_\perp^2}\Big\{
{\Gamma(\epsilon)\over p_\perp^{2\epsilon}},\partial_\perp^2U^{ab}\Big\}
{1\over p_\perp^2}|y_\perp)\Big]
+{1\over 3\epsilon}
\Big[2 (x_\perp|{1\over p_\perp^2}\partial_\perp^2U^{ab}{1\over p_\perp^2}|y_\perp)
-(x_\perp|{1\over p_\perp^2}\partial_\perp^2U^{ab}{1\over p_\perp^2}|x_\perp)
-(y_\perp|{1\over p_\perp^2}\partial_\perp^2U^{ab}{1\over p_\perp^2}|y_\perp)\Big]
\Bigg)
\nonumber
\end{eqnarray}
where the last term $\sim{1\over 3\epsilon}$ is a counterterm calculated in the Appendix.

\subsection{Quark loop inside the shock wave}
\begin{figure}
\includegraphics[width=0.23\textwidth]{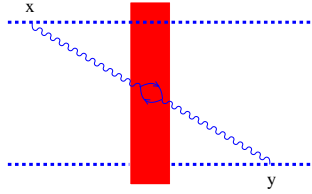}
\caption{Quark loop inside the shock wave \label{quarkloop2}.}
\end{figure}

\begin{figure}
\includegraphics[width=0.95\textwidth]{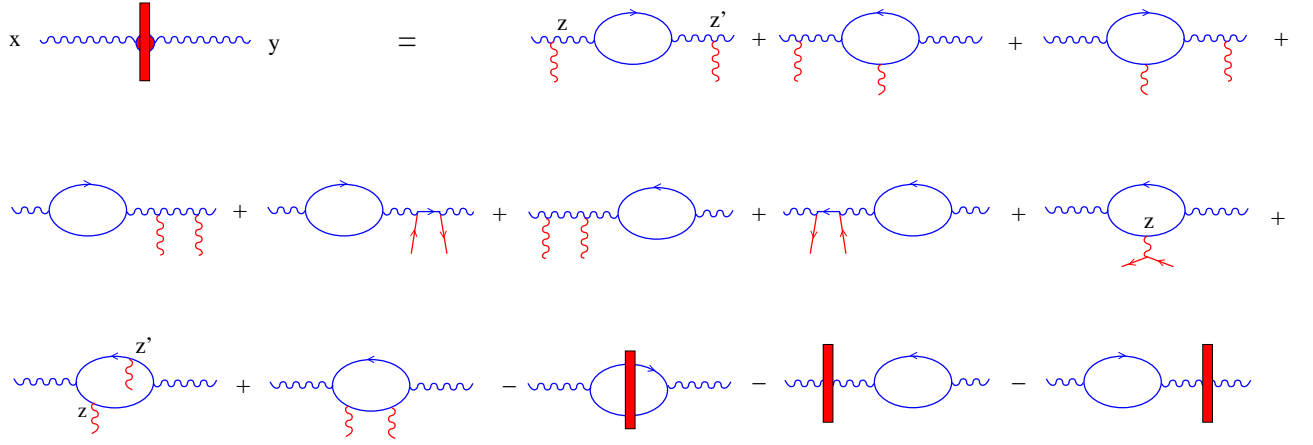}
\caption{A possible local contibution coming from the quark loop inside the shock wave \label{kvnlocal}.}
\end{figure}

Let us consider the diagram in Fig. \ref{quarkloop2} with the quark loop inside the shock wave of width $\lambda=\Delta z_\ast\sim {s\over m^2}e^{-\eta_2}$ where $m^2$ 
is some characteristic mass scale of order of $Q^2$.  
From the form of the perturbative propagator 
$$
(z|{1\over p^2+i\epsilon}|z')={s\over 8\pi x_\ast}\!\int_0^\infty \!{\dhd\alpha\over\alpha}
e^{-i\alpha(z-z')_\bullet-i{(z-z')_\perp^2\over 4(z-z')_\ast}\alpha s}
$$
we see that the  characteristic transverse scale inside the shock wave is  
\begin{equation}
(z-z')_\perp^2\sim {\lambda\over\alpha s}\simeq m^{-2}{e^{-\eta_2}\over\alpha}\ll m^{-2}
\end{equation}
and therefore the contribution of the diagram in Fig. \ref{quarkloop2} reduces
to the contribution of some operator {\it local} in the transverse space. 
By dimensional arguments,  this local operator must have the same twist as 
the operator describing the interaction of the gluon with the shock wave at the tree level.
In the leading order in $\alpha_s$, the vertex of interaction of gluon with the shock-wave field is proportional to 
\begin{eqnarray}
&&\partial_\perp^2U~=~-ig[DG]+2g^2[GG]
\label{dkvau}
\end{eqnarray}
where
\begin{eqnarray}
&&\hspace{-6mm}[DG]\equiv \int\! du~ [\infty p_1,up_1]_xD^iG_{i\bullet}(up_1+x_\perp)[up_1,-\infty p_1]_x
\label{dgigg}\\
&&\hspace{-6mm}[GG]\equiv \int\! du dv ~\theta(u-v)~[\infty p_1,up_1]_xG_{\bullet i}(up_1+x_\perp)[up_1,vp_1]_x
G_\bullet^{~i}(vp_1+x_\perp)[vp_1,-\infty p_1]_x
\nonumber
\end{eqnarray}
These operators have twist 2 so a possible local operator describing the
 gluon interaction with the shock wave at the one-loop level must also be of twist 2.
To find this local operator, we 
consider (the quark loop contribution to) the  color dipole
 ${\rm tr}\{U_x U^\dagger_y\}$ at small 
 at $(x-y)_\perp^2\rightarrow 0$ and compare the expansion of the 
 contribution of the diagrams in Fig. \ref{kvloop1} to the exact 
 calculation of the light-cone expansion of ${\rm tr}\{U_x U^\dagger_y\}$ in QCD (up to twist-2 level), see Fig. \ref{kvnlocal}.

The first step is the light-cone expansion of the sum of the diagrams in the  in Figs. \ref{kvloop1}- \ref{kvloop4} in the shock-wave background. 
The light-cone expansion of Eq. (\ref{vesvklad1}) at $x_\perp\rightarrow y_\perp$ starts 
with 
terms quadratic in $q$($q'$). They lead to the  operators $\partial^2U_x$ and $\partial_iU_x\partial_i U^\dagger_x$ (in the leading order we do not distinguish between
$U_x$ and $U_y$):
\begin{eqnarray}
&&\hspace{-6mm}
{\rm Tr}\{\partial_iU_x\partial_iU^\dagger_y\} \stackrel{x\rightarrow y_\perp}{=}
{\alpha^2_s\over \pi}n_f
\Delta\eta
U_x^{ab}~\Bigg[
{4\over 3}\mu^{2-d} 
B(1-\epsilon,1-\epsilon)
(x_\perp|\Big([3-{1\over\epsilon}]\delta_{ij}-{p_ip_j\over p_\perp^2}\Big)
{\Gamma(1+\epsilon)\over (p_\perp^2)^{1+\epsilon}}|y_\perp)
{\rm Tr}\{t^a\partial_i U_xt^b\partial_jU^\dagger_x\}
\nonumber\\
&&\hspace{-6mm}
+~
4\mu^{2-d} B(2-\epsilon,2-\epsilon)
(x_\perp|
{\Gamma(\epsilon)\over (p_\perp^2)^{1+\epsilon}}|y_\perp)
\partial_\perp^2U_x^{ab}
-{1\over 3\epsilon}
(x_\perp|{1\over p_\perp^2}|y_\perp)\partial_\perp^2U_x^{ab}
\Bigg]
\nonumber\\
&&\hspace{-6mm}
=~
{\alpha^2_s\over \pi^2}n_f
\Delta\eta
~\Bigg\{
{1\over 6N_c}\mu^{2\epsilon} 
B(1-\epsilon,1-\epsilon)
\Big[\Big({1\over\epsilon}-3+{1\over 2(1+\epsilon)}\Big)\delta_{ij}
+{2\epsilon\over (1+\epsilon)}{\Delta_i\Delta_j\over \Delta_\perp^2}\Big]
{\Gamma(-2\epsilon)\over (\Delta_\perp^2)^{-2\epsilon}}
{\rm Tr}\{\partial_i U_x\partial_jU^\dagger_x\}
\nonumber\\
&&\hspace{-6mm}
+~
{\mu^{2\epsilon}\over\epsilon} B(2-\epsilon,2-\epsilon)
{\Gamma(-2\epsilon)\over (\Delta_\perp^2)^{-2\epsilon}}
-{1\over 12\epsilon}{\Gamma(-\epsilon)\over (\Delta_\perp^2)^{-\epsilon}}\Big]
U_x^{an}\partial_\perp^2U_x^{ab}
\Bigg\}
\label{vesvkladlikone}
\end{eqnarray}
The expression (\ref{vesvkladlikone}) should be compared to the light-cone expansion
of the quark-loop part of the gluon propagator in an external field (see Fig. \ref{kvnlocal})
performed in the Appendix.
A typical term of the light-cone expansion has the form:
\begin{eqnarray}
&&\hspace{-0mm}
A^a_\bullet(x_\ast,x_\perp)A^b_\bullet(y_\ast,y_\perp)~
=~{ig^3\over 16\pi^2}B(2-\epsilon,2-\epsilon)\!\int_0^\infty\! {\dhd\alpha\over \alpha^3}\Big({4i\over\alpha s}\Big)^\epsilon
 \Big(-{i\alpha s\over 4\pi\Delta_\ast}\Big)^{1-\epsilon}
e^{i{\Delta_\perp^2\over 4\Delta_\ast}\alpha s}
\label{tipikal}\\
&&\hspace{-6mm}
\times~\Bigg[-ig
\!\int_{y_\ast}^{x_\ast}\!\!
d{2\over s}z_\ast ~{1\over\epsilon}[(x-z)_\ast^\epsilon+ (z-y)_\ast^\epsilon]~
([x_\ast,z_\ast]_xD^iG_{i\bu}(z_\ast,x_\perp)[z_\ast,z'_\ast]_x
[z_\ast,y_\ast]_x)^{ab}
\nonumber\\
&&\hspace{-6mm}
+~g62\!\int_{y_\ast}^{x_\ast}\!\!
d{2\over s}z_\ast \!\int_{y_\ast}^{z_\ast}\!\! d{2\over s}z'_\ast~
([x_\ast,z_\ast]_xG_\bu^{~i}(z_\ast,x_\perp)[z_\ast,z'_\ast]_x
G_{\bu i}(z'_\ast,x_\perp)[z'_\ast,y_\ast]_x)^{ab}
\Big\{
 {1\over\epsilon}[(x-z)_\ast^\epsilon
+ (z'-y)_\ast^\epsilon]+(z-z')_\ast^{\epsilon}
 \Big\}\Bigg]
\nonumber
\end{eqnarray}
In our ``external'' field the characteristic distances $z_\ast(z'_\ast)$ are of the order of width of the shock wave: $z_\ast,z'_\ast\sim e^{\eta_2}\sqrt{s/m^2}$. As we shall see below, the characteristic distances $x_\ast$ and $y_\ast$ are 
$\sim e^{\eta_1}\sqrt{s/m^2}$
so we can neglect $z_\ast$ and $z'_\ast$ in comparison to  $x_\ast$ and/or $y_\ast$.
The formula (\ref{vkladot1}) simplifies to
\begin{eqnarray}
&&\hspace{-0mm}
A^a_\bullet(x_\ast,x_\perp)A^b_\bullet(y_\ast,y_\perp)~
=~{ig^3\over 16\pi^2}B(2-\epsilon,2-\epsilon)\!\int_0^\infty\! {\dhd\alpha\over \alpha^3}\Big({4i\over\alpha s}\Big)^\epsilon
 \Big(-{i\alpha s\over 4\pi\Delta_\ast}\Big)^{1-\epsilon}
e^{i{\Delta_\perp^2\over 4\Delta_\ast}\alpha s}
\label{tipikalvklad}\\
&&\hspace{-6mm}
\times~\Bigg[-ig{1\over\epsilon}
\!\int_{y_\ast}^{x_\ast}\!\!
d{2\over s}z_\ast ~[x_\ast^\epsilon+ (-y)_\ast^\epsilon]
([x_\ast,z_\ast]_xD^iG_{i\bu}(z_\ast,x_\perp)[z_\ast,z'_\ast]_x
[z_\ast,y_\ast]_x)^{ab}
\nonumber\\
&&\hspace{-6mm}
+~g^2\!\int_{y_\ast}^{x_\ast}\!\!
d{2\over s}z_\ast \!\int_{y_\ast}^{z_\ast}\!\! d{2\over s}z'_\ast~
([x_\ast,z_\ast]_xG_\bu^{~i}(z_\ast,x_\perp)[z_\ast,z'_\ast]_x
G_{\bu i}(z'_\ast,x_\perp)[z'_\ast,y_\ast]_x)^{ab}
\Big\{
 {1\over\epsilon}[x_\ast^\epsilon
+ (-y)_\ast^\epsilon]+(z-z')_\ast^{\epsilon}
 \Big\}\Bigg]
\nonumber
\end{eqnarray}
A very important observation is that 
the contributions proportional to
\begin{equation}
g^4n_f \int\!dz_\ast\!\int dz'_\ast~\theta(z-z')~(z-z')^\epsilon G_{\bu i}(z_\ast)G_{\bu i}(z'_\ast)
\label{badterm}
\end{equation}
present in the individual diagrams in Fig. \ref{kvnlo14}, cancel their sum.  If it were not true, there would be an addtional contribution to the gluon propagator (\ref{gluprop}) at the $g^4$ level coming from the small-size (large-momenta) 
quark loop. Indeed, the calculations of Feynman diagrams with the propagators (\ref{gluprop}) and (\ref{kvpropagator}) imply that we first take limit $z_\ast,z'_\ast\rightarrow 0$ and limit $d_\perp\rightarrow 2$ afterwards. 
With such order of limits, the contribution (\ref{badterm}) vanishes. However, 
the proper order of these limits is to first take $d_\perp\rightarrow 2$ (which 
will give finite expressions after adding the counterterms) and then try to impose 
condition that the external field is very narrow by taking the limit $z_\ast, z'_\ast\rightarrow 0$. In this case, Eq. (\ref{badterm}) reduces to 
$g^4n_f[GG]$.
 The non-commutativity of these limits would mean that 
the contribution ${1\over p^2}[GG]{1\over p^2}$ should be added to the gluon propagator (\ref{gluprop}) to restore the correct result. Fortunately, the terms $\sim$
(\ref{badterm}) cancel which means that there are no additional contributions to the 
gluon propagator coming from the quark loop inside the shock wave ($\equiv$ quark loop
with large momenta).
 
Since there is no external field outside the shock wave,  after cancellation of the terms $\sim (z-z')^\epsilon$ we see that at $x_\ast y_\ast>0$  
 Eq. (\ref{tipikalvklad}) vanishes, and at  $x_\ast>0>y_\ast$ one can extend the limits of integration in the gluon operators
to $\pm\infty$ and obtain 
\begin{eqnarray}
&&\hspace{-0mm}
A^a_\bullet(x_\ast,x_\perp)A^b_\bullet(y_\ast,y_\perp)~
=~{ig^2\over 16\pi^2}B(2-\epsilon,2-\epsilon)\!\int_0^\infty\! {\dhd\alpha\over \alpha^3}\Big({4i\over\alpha s}\Big)^\epsilon
 \Big(-{i\alpha s\over 4\pi\Delta_\ast}\Big)^{1-\epsilon}
e^{i{\Delta_\perp^2\over 4\Delta_\ast}\alpha s}
\label{tipikalwilson}\\
&&\hspace{-6mm}
\times~\Bigg[-ig
{1\over\epsilon}[x_\ast^\epsilon+ (-y)_\ast^\epsilon]
[DG]^{ab}
+~g^2[GG]^{ab}
 {1\over\epsilon}[x_\ast^\epsilon
+ (-y)_\ast^\epsilon]\Bigg]
\nonumber\\
&&\hspace{-6mm}
=~{ig^2\over 16\pi^2}B(2-\epsilon,2-\epsilon)\!\int_0^\infty\! {\dhd\alpha\over \alpha^3}\Big({4i\over\alpha s}\Big)^\epsilon
 \Big(-{i\alpha s\over 4\pi\Delta_\ast}\Big)^{1-\epsilon}
e^{i{\Delta_\perp^2\over 4\Delta_\ast}\alpha s}
 {1\over\epsilon}[x_\ast^\epsilon
+ (-y)_\ast^\epsilon]~\partial_\perp^2U^{ab}_x
\nonumber
\end{eqnarray}
The light-cone expansion  of gluon propagator contains only Wilson lines
and their derivatives as should be expected after cancellation of the ``contaminating''
terms (\ref{badterm}).

We have demonstrated in the Appendix that the light-cone expansion of the quark-loop contribution to the gluon propagator coincides with Eq. (\ref{vesvkladlikone}) as should be expected once we established the commutativity of the limits $d_\perp\rightarrow 2$ and
 $z_\ast(z'_\ast)\rightarrow 0$.

\subsection{Quark loop in the coordinate representation}
To calculate the integrals over momenta in Eq. (\ref{vesvklad1}) 
it is convenient to subtract (and add) ${\rm Tr}\{t^aU_zt^bU^\dagger_z\}$ from
${\rm Tr}\{t^aU_zt^bU^\dagger_{z'}\}$: 
\begin{equation}
{\rm Tr}\{t^aU_zt^bU^\dagger_{z'}\}={\rm Tr}\{t^aU_zt^bU^\dagger_{z'}-t^aU_zt^bU^\dagger_z\}+{\rm Tr}\{t^aU_zt^bU^\dagger_z\}
\label{adunsybtract}
\end{equation}
Let us start with the last term in the r.h.s. of Eq. (\ref{adunsybtract}).
In the momentum representation, this term corresponds to
${\rm Tr}\{t^a
U(q)t^b U^{\dagger}(q')\}\rightarrow
4\pi^2 \delta(q')\!\int\! dz e^{i(q,z)_\perp}{\rm Tr}\{t^aU_zt^bU^\dagger_z\}=
2\pi^2 \delta(q')U^{ab}(q)$
so we get
\begin{eqnarray}
&&\hspace{-16mm}
-2\alpha^2_sn_f
\Delta\eta
t^aU_x\otimes t^bU^{\dagger}_y
{\mu^{2\epsilon}\over \pi}
\int\! \dhd^dp~\dhd^dq ~{e^{i(p,x)_\perp-i(p-q,y)_\perp}\over p^2(p-q)^2}
U^{ab}(q)\!\int_0^1 \! dv du~
\Bigg[-
{\bar{u}uq^2\Gamma(\epsilon)\over 
[(p-qu)^2\bar{v} v+q^2\bar{v} \bar{u} u]^\epsilon}
\nonumber\\
&&\hspace{-16mm}
+~{q^2\bar{u}u\bar{v}\Gamma(1+\epsilon)\over 
[(p-qu)^2\bar{v} v+q^2\bar{v} \bar{u} u]^{1+\epsilon}}
\Big\{(p-qu)^2(-1+2v)
+\bar{u}u q^2\Big\}\Bigg]
\nonumber\\
&&\hspace{-16mm}
=~2{\alpha^2_s\over\pi}n_f
\Delta\eta
t^aU_x\otimes t^bU^{\dagger}_y
\int\! \dhd^dp~\dhd^dq ~{e^{i(p,x)_\perp-i(p-q,y)_\perp} \over p^2(p-q)^2}
U^{ab}(q)~q^2
{\Gamma(\epsilon)\over
(q^2/\mu^2)^\epsilon}B(2-\epsilon,2-\epsilon)
\label{vklad3}
\end{eqnarray}
where we have used integration by parts to transform the second term in the l.h.s of this equation. Alternatively, this result can be obtained directly from Eq. (\ref{kvlup1a}) 
after the substitution (\ref{adunsybtract}).

 For future use we need to rewrite it in Schwinger's representation:
\begin{eqnarray}
-2 {\alpha^2_s\over\pi}\mu^{2-d}n_f
\Delta\eta
\{t^aU_x\otimes t^bU^{\dagger}_y\}
(x|{1\over p^2}
\Big({\Gamma(\epsilon)\over
(-\partial^2)^\epsilon}\partial_\perp^2U^{ab}\Big){1\over p^2}|y)
B(2-\epsilon,2-\epsilon)
\label{uvklad}
\end{eqnarray}
The contribution coming from the first term in Eq. (\ref{adunsybtract}) is UV-finite. To calculate
it in coordinate representation it is convenient to return back to the original expression
\begin{eqnarray}
&&\hspace{-26mm}
-8\alpha^2_sn_f
\Delta\eta
t^aU_x\otimes t^bU^{\dagger}_y
\int \dhd^2k_1\dhd^2k'_1\dhd^2k_2\dhd^2k'_2~e^{i(k_1,x-z)_\perp+i(k'_1,x-z')_{\perp}
-i(k_,y-z)_\perp-i(k'_2,y-z')_{\perp}}
\nonumber\\
&&\hspace{-1mm}
{\rm Tr}\{t^aU_zt^bU^\dagger_{z'}-t^aU_zt^bU^\dagger_z\}
\int_0^1 du~
{(k_1,k_2)(k'_1,k'_2)
+(k_1,k'_1)(k_2,k'_2)-(k_1,k'_2)(k'_1,k_2)
\over (k_1+k'_1)^2(k_2+k'_2)^2
(k^2_1u+{k'_1}^2\bar{u})(k^2_2u+{k'_2}^2\bar{u})}     
\nonumber
\end{eqnarray}
and use the formulas
\begin{eqnarray}
&&\hspace{-25mm}
\int\dhd^d k_1\dhd^d k_2~{(k_1,k_2)e^{i(k_1,x_1)+i(k_2,x_2)}
\over(k_1+k_2)^2(k_1^2\bar{u}+k_2^2u)}~=~
{1\over 4\pi^2(x_1-x_2)^2}\Big[1-{(x_1,x_2)\over x_1^2 u+x_2^2\bar{u}}\Big]
+O(d-2)
\nonumber\\
&&\hspace{-25mm}
\int\dhd^d k_1\dhd^d k_2~{k_{1i}k_{2j}-i\leftrightarrow j
\over(k_1+k_2)^2(k_1^2\bar{u}+k_2^2u)}~e^{i(k_1,x_1)+i(k_2,x_2)}~=~-
{x_{1i}x_{2j}-i\leftrightarrow j\over 4\pi^2(x_1-x_2)^2(x_1^2 u+x_2^2\bar{u})}
+O(d-2)
\label{formula}
\end{eqnarray}

 After some algebra, one obtains:
\begin{eqnarray}
&&\hspace{-26mm}U_x\otimes U^{\dagger}_y~=~
-{\alpha^2_s\over 2\pi^4}
\Delta\eta\{t^aU_x\otimes t^bU^{\dagger}_y\}
\!\int\! d^2z d^2z' \!\int_0^1\! du~
{\rm Tr}\{t^aU_zt^b U^\dagger_{z'}-t^aU_zt^b U^\dagger_z\}~{1\over (z-z')^4}
\nonumber\\
&&\hspace{-6mm}
\times~
\Bigg[1-{(X,X')\over X^2u+{X'}^2\bar{u}}-{(Y,Y')\over Y^2u+{Y'}^2\bar{u}}
+{(X,X')(Y,Y')+(X,Y)(X'Y')-(X,Y')(X',Y)\over (X^2u+{X'}^2\bar{u})(Y^2u+{Y'}^2\bar{u})}
\label{otvet1}
\end{eqnarray}
where $X\equiv x-z$, $X'\equiv x-z'$, $Y\equiv y-z$, $y'\equiv y-z'$. Note that the
singularity at $z'= z$ is integrable.

Performing the integation over $u$ and adding the $UV$-divergent term
(\ref{uvklad}) one obtains the total contribution of the diagram in Fig. \ref{kvloop1}a in the form:
\begin{eqnarray}
&&\hspace{-16mm}(U_x\otimes U^{\dagger}_y)^{\rm Fig.\ref{kvloop1}a}
~=~-{\alpha^2_s\over 2\pi^4}
\Delta\eta\{t^aU_x\otimes t^bU^{\dagger}_y\}
\!\int\!d^2z d^2z'~
{\rm Tr}\{t^a
U_zt^b U^\dagger_{z'}-t^aU_zt^bU^\dagger_z\}~{1\over (z-z')^4}
\nonumber\\
&&\hspace{-16mm}
\times~
\Bigg\{1-{(X,X')\over X^2-{X'}^2}\ln{X^2\over{X'}^2}-
{(Y,Y')\over Y^2-{Y'}^2}\ln{Y^2\over {Y'}^2}
+{(X,X')(Y,Y')+(X,Y)(X'Y')-(X,Y')(X',Y)\over {X'}^2Y^2-{Y'}^2X^2}
\ln{{X'}^2Y^2\over {Y'}^2X^2}\Bigg\}
\nonumber\\
&&\hspace{-16mm}
-~
2{\alpha^2_s\over\pi}\mu^{2\epsilon}B(2-\epsilon,2-\epsilon)
\Delta\eta
\{t^aU_x\otimes t^bU^{\dagger}_y\}
(x_\perp|{1\over p_\perp^2}\Big(
{\Gamma(\epsilon)\over (-\partial_\perp^2)^\epsilon}
\partial_\perp^2U\Big)^{ab}{1\over p_\perp^2}|y_\perp)
\label{otvet2}
\end{eqnarray}

Sum of the UV-divergent contributions takes the form
\begin{eqnarray}
&&\hspace{-16mm}
2n_f\alpha^2_s{\mu^{2\epsilon}\over \pi} B(2-\epsilon,2-\epsilon)
\Delta\eta
t^aU_x\otimes t^bU^{\dagger}_y
(x_\perp|{1\over p_\perp^2}\Big\{
{\Gamma(\epsilon\over (p_\perp^2)^\epsilon},\partial_\perp^2U\Big\}
{1\over p_\perp^2}
-{1\over p_\perp^2}\Big(
{\Gamma(\epsilon)\over (-\partial_\perp^2)^\epsilon}
\partial_\perp^2U\Big){1\over p_\perp^2}
|y_\perp)^{ab}
\label{sumuv1}
\end{eqnarray}

The counterterm  is 
calculated in the Appendix (we use the $\overline{MS}$ scheme):
\begin{eqnarray}
&&\hspace{-16mm}
-2n_f\alpha^2_s 
\Delta\eta
t^aU_x\otimes t^bU^{\dagger}_y
{1\over 3\pi\epsilon}
(x_\perp|{1\over p_\perp^2}\partial_\perp^2U^{ab}{1\over p_\perp^2}|y_\perp)
\label{konterm}
\end{eqnarray}
Adding the counterterm,  one gets after some algebra 
\begin{eqnarray}
&&\hspace{-16mm}
n_f{\alpha^2_s\over 3\pi} 
\Delta\eta
t^aU_x\otimes t^bU^{\dagger}_y
(x_\perp|{1\over p_\perp^2}\Big[\Big\{\ln{\mu^2\over p_\perp^2},
\partial_\perp^2U\Big\}-\Big(\ln{\mu^2\over -\partial_\perp^2}\partial_\perp^2U\Big)
+{5\over 3}\partial_\perp^2U\Big]{1\over p_\perp^2}|y_\perp)^{ab}
\nonumber\\
&&\hspace{-16mm}
=~-n_f{\alpha^2_s\over 3\pi} 
\Delta\eta
t^aU_x\otimes t^bU^{\dagger}_y
\int\! {d^2z\over 8\pi^2}~\Big\{
{(x-y)^2\over X^2Y^2}~
[\ln(x-y)^2\mu^2+{5\over 3}]
\nonumber\\
&&\hspace{-16mm}
-~{1\over X^2}[\ln Y^2\mu^2+{5\over 3}]
-{1\over Y^2}[\ln X^2\mu^2+{5\over 3}]
\Big\}(2U_z-U_x-U_y)^{ab}
\label{sumuv1}
\end{eqnarray}
The calculation is simplified if one notes that $\partial_\perp^2U_z$ in the l.h.s. 
can be repaced by $\partial_\perp^2(U_z-{1\over 2}U_x-{1\over 2}U_y)$.
Our final result for the sum of the diagrams in Fig. \ref{kvloop1} 
has the form
\begin{eqnarray}
&&\hspace{-6mm}\Big(U_x\otimes U^{\dagger}_y\Big)_{\rm Fig. \ref{kvloop1}}~=~-{\alpha^2_s\over 2\pi^3}
\Delta\eta~n_f
\{t^aU_x\otimes t^bU^{\dagger}_y\}
\!\int\!dz\Bigg[{1\over \pi}\!\int\! dz'~
{\rm Tr}^{\rm col}\{t^a
U_zt^b U^\dagger_{z'}-t^aU_zt^bU^\dagger_z\}~{1\over (z-z')^4}
\nonumber\\
&&\hspace{-6mm}
\times~
\Bigg\{1-{(X,X')\over X^2-{X'}^2}\ln{X^2\over{X'}^2}-
{(Y,Y')\over Y^2-{Y'}^2}\ln{Y^2\over {Y'}^2}
+{(X,X')(Y,Y')+(X,Y)(X'Y')-(X,Y')(X',Y)\over {X'}^2Y^2-{Y'}^2X^2}
\ln{{X'}^2Y^2\over {Y'}^2X^2}\Bigg\}
\nonumber\\
&&\hspace{-6mm}
+~{1\over 12} \Big\{
{(x-y)^2\over X^2Y^2}~
[\ln(x-y)^2\mu^2+{5\over 3}]
-{1\over X^2}[\ln Y^2\mu^2+{5\over 3}]
-{1\over Y^2}[\ln X^2\mu^2+{5\over 3}]
\Big\}(2U_z-U_x-U_y)^{ab}\Bigg]
\label{otvet3}
\end{eqnarray}
As we mentioned above, 
the contribution of diagrams in Fig. \ref{kvloop2} is obtained from
Eq. (\ref{otvet3}) by replacement
 $t^aU_x\otimes t^bU^{\dagger}_y\rightarrow U_xt^b\otimes U^\dagger_y t^a$ and
the contribution of the diagram in Fig. \ref{kvloop3} can be obtained from Eq. (\ref{otvet3}) by taking $y=x$ in the integrand (and changing the sign)
\begin{eqnarray}
&&\hspace{-6mm}
\Big(U_x\otimes U^{\dagger}_y\Big)_{\rm Fig. \ref{kvloop3}}~=~{\alpha^2_s\over \pi^3}
\Delta\eta~n_f
\{t^aU_x t^b\otimes U^{\dagger}_y\}
\!\int\!dz\Bigg[{1\over \pi}\!\int\! dz'~
{\rm Tr}\{t^a
U_zt^b U^\dagger_{z'}-t^aU_zt^bU^\dagger_z\}~{1\over (z-z')^4}
\nonumber\\
&&\hspace{-6mm}
\times~
\Big(1-{(X,X')\over X^2-{X'}^2}\ln{X^2\over{X'}^2}\Big)
- {1\over 6X^2}[\ln X^2\mu^2+{5\over 3}](U_z-U_x)^{ab}\Bigg]
\label{regg1}
\end{eqnarray}
Similarly, the contribution of the diagram in Fig. \ref{kvloop4} is obtained by the replacement $x\rightarrow y$:
\begin{eqnarray}
&&\hspace{-6mm}
\Big(U_x\otimes U^{\dagger}_y\Big)_{\rm Fig. \ref{kvloop4}}~=~{\alpha^2_s\over \pi^3}
\Delta\eta~n_f
\{U_x \otimes t^bU^{\dagger}_yt^a\}
\!\int\!dz\Bigg[{1\over \pi}\!\int\! dz'~
{\rm Tr}^{\rm col}\{t^a
U_zt^b U^\dagger_{z'}-t^aU_zt^bU^\dagger_z\}~{1\over (z-z')^4}
\nonumber\\
&&\hspace{-6mm}
\times~
\Big(1-{(Y,Y')\over Y^2-{Y'}^2}\ln{Y^2\over{Y'}^2}\Big)
- {1\over 6Y^2}[\ln Y^2\mu^2+{5\over 3}](U_z-U_y)^{ab}\Bigg]
\label{regg2}
\end{eqnarray}

Summing the contributions of the diagrams in Fig. \ref{kvloop1} - \ref{kvloop4} and taking Tr over the color indices, one obtains
\begin{eqnarray}
&&\hspace{-6mm}{\rm Tr}\{U_x U^{\dagger}_y\}~=~{\alpha^2_s\over \pi^3}
\Delta\eta~n_f
{\rm Tr}\{t^aU_xt^bU^{\dagger}_y\}
\!\int\!d^2z
\nonumber\\
&&\hspace{-6mm}
\times~
\Bigg[{1\over \pi}\!\int\! d^2z'~
{\rm Tr}\{t^a
U_zt^b U^\dagger_{z'}-t^aU_zt^bU^\dagger_z\}~{1\over (z-z')^4}
\Bigg\{1
-{{X'}^2Y^2+{Y'}^2X^2-(x-y)^2(z-z')^2\over 2({X'}^2Y^2-{Y'}^2X^2)}
\ln{{X'}^2Y^2\over {Y'}^2X^2}\Bigg\}
\nonumber\\
&&\hspace{-6mm}
+~{1\over 12} \Big\{
-{(x-y)^2\over X^2Y^2}~
[\ln(x-y)^2\mu^2+{5\over 3}]
+{X^2-Y^2\over X^2Y^2}\ln{X^2\over Y^2}
\Big\}(2U_z-U_x-U_y)^{ab}\Bigg]
\label{otvetrace}
\end{eqnarray}

Let us present the total result for the sum of the leading order BK equation and the quark NLO correction 
\begin{eqnarray}
&&\hspace{-6mm}
{d\over d\eta}{\rm Tr}\{U_x U^{\dagger}_y\}~=~{\alpha_s\over 2\pi^2}
\!\int\!d^2z~[{\rm Tr}\{U_x U^{\dagger}_z\}{\rm Tr}\{U_z U^{\dagger}_y\}
-N_c{\rm Tr}\{U_x U^{\dagger}_y\}]
\nonumber\\
&&\hspace{6mm}
\times~
\Big[{(x-y)^2\over X^2 Y^2}
\Big(1-{\alpha_s n_f\over 6\pi}[\ln (x-y)^2\mu^2+{5\over 3}]\Big)+
{\alpha_s n_f\over 6\pi}{X^2-Y^2\over X^2Y^2}\ln{X^2\over Y^2}\Big]
\nonumber\\
&&\hspace{-6mm}
~+{\alpha^2_s\over \pi^4}n_f
{\rm Tr}\{t^aU_xt^bU^{\dagger}_y\}
\!\int\!d^2z d^2z'~
{\rm Tr}\{t^a
U_zt^b U^\dagger_{z'}-t^aU_zt^bU^\dagger_z\}~{1\over (z-z')^4}
\nonumber\\
&&\hspace{6mm}
\times~
\Big\{1
-{{X'}^2Y^2+{Y'}^2X^2-(x-y)^2(z-z')^2\over 2({X'}^2Y^2-{Y'}^2X^2)}
\ln{{X'}^2Y^2\over {Y'}^2X^2}\Big\}
\label{nlobk}
\end{eqnarray}

We see the first  term proportional to $\ln(...)^2\mu^2$ (we will call it the ``UV'' term) has the same structure
as the zero-order contribution (\ref{bk}). In the next Section we will use it to determine the argument of the running coupling constant in Eq. (\ref{bk}).

\subsection{Comparison to NLO BFKL}

To compare with the NLO BFKL equation we need to linearize Eq. (\ref{nlobk}) which gives
\begin{eqnarray}
&&\hspace{-6mm}
{d\over d\eta}{\cal U}(x,y)~=~{\alpha_sN_c\over 2\pi^2}
\!\int\!d^2z~({\cal U}(x,z)+{\cal U}(z,y)-{\cal U}(x,y)]
\nonumber\\
&&\hspace{6mm}
\times~
\Big[{(x-y)^2\over X^2 Y^2}
\Big(1-{\alpha_s n_f\over 6\pi}[\ln (x-y)^2\mu^2+{5\over 3}]\Big)+
{\alpha_s n_f\over 6\pi}{X^2-Y^2\over X^2Y^2}\ln{X^2\over Y^2}\Big]
\nonumber\\
&&\hspace{-6mm}
~-{\alpha^2_sn_f\over 4N_c\pi^4}
\!\int\!d^2z d^2z'~
{{\cal U}(z,z')\over (z-z')^4}
\Big\{1
-{{X'}^2Y^2+{Y'}^2X^2-(x-y)^2(z-z')^2\over 2({X'}^2Y^2-{Y'}^2X^2)}
\ln{{X'}^2Y^2\over {Y'}^2X^2}\Big\}
\label{nlobklin}
\end{eqnarray}
This should be compared to the quark part of the non-forward BFKL 
kernel \cite{fadin} but the Fourier
transformation from the momentum space to the dipole-type representation appears to be rather difficult. 

To simplify the comparison, let us consider the case of forward scattering and 
write down the Mellin representation of ${\cal U}(x,y)$
\begin{eqnarray}
&&\hspace{-6mm}
{\cal U}^\eta(x-y)~=~\int\! d\nu~ (x-y)^{2\gamma}~{\cal U}^\eta_\nu,~~~~~
\gamma\equiv {1\over 2}+i\nu
\label{mellin}
\end{eqnarray}
where we have displayed the dependence on the rapidity $\eta$ explicitly. 
Using the integrals ($\chi(\gamma)\equiv [-\psi(\gamma)-\psi(1-\gamma)+2\psi(1)]$)
\begin{eqnarray}
&&\hspace{-6mm}
\int\! d^2z ~{(x-y)^2\over X^2Y^2}
\Big[(X^2)^\gamma+(Y^2)^\gamma
-((x-y)^2)^\gamma\Big]~=~2\pi \chi(\gamma)((x-y)^2)^\gamma
\label{ints}\\
&&\hspace{-6mm}
\int\! d^2z \Big[{\ln X^2/Y^2\over Y^2}-{\ln X^2/Y^2\over X^2}\Big]
(X^{2\gamma}+Y^{2\gamma}-(x-y)^{2\gamma})
~=~\pi(x-y)^{2\gamma}\{\psi'(1-\gamma)-\psi'(\gamma)-\chi^2(\gamma) +{4\over\gamma}\chi(\gamma)\}
\nonumber\\
&&\hspace{-6mm}
\!\int\!{d^2z d^2z'\over (z-z')^{4-2\gamma}}~
\Big\{1
-{{X'}^2Y^2+{Y'}^2X^2-(x-y)^2(z-z')^2\over 2({X'}^2Y^2-{Y'}^2X^2)}
\ln{{X'}^2Y^2\over {Y'}^2X^2}\Big\}~
=~{\pi^4\cos\pi\gamma\over\sin^2\pi\gamma}{2+3\gamma(1-\gamma)
\over (1-4\gamma^2)(3-2\gamma)}
(x-y)^{2\gamma}
\nonumber
\end{eqnarray}
we obtain
\begin{eqnarray}
&&\hspace{-6mm}
{d\over d\eta}~{\cal U}^\eta_\nu~=~
{\alpha_sN_c\over \pi}[1-{\alpha_s\over 6\pi}n_f\ln(x-y)^2\mu^2]
\label{nlobfkl}\\
&&\hspace{6mm}
\times~
\Big(\chi(\gamma)-{\alpha_s\over 12\pi}n_f\Big\{[\psi'(\gamma)-\psi'(1-\gamma)+\chi^2(\gamma)-{4\over\gamma}\chi(\gamma)]
+{10\over 3}\chi(\gamma)
+{3\pi^2\cos\pi\gamma\over N_c^2\sin^2\pi\gamma}{2+3\gamma(1-\gamma)\over (1-4\gamma^2)(3-2\gamma)}
\Big\}\Big)~{\cal U}^\eta_\nu
\nonumber
\end{eqnarray}
where $\gamma={1\over 2}+i\nu$.  This expression should be be compared to the NLO BFKL result \cite{nlobfkl}. Unfortunately, there is no explicit expression for the coordinate-space NLO BFKL kernel yet.  However, the last two terms in braces in r.h.s. of this Equation coincide with the expression for the $n_f$ part of the  eigenvalue $\delta(\gamma)$ of Ref. \onlinecite{nlobfkl}. 
The first term in braces should correspond to the quark part of $\beta$-function
contribution to the eigenvalue $\delta(\gamma)$.
We expect to study the relation to NLO BFKL in detail after completing the calculation of the gluon loop.

\section{Bubble chain and the argument of coupling constant.}

\begin{figure}
\includegraphics[width=0.6\textwidth]{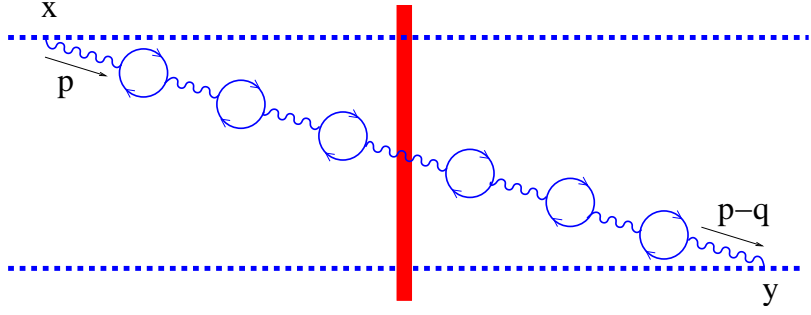}
\caption{Bubble chain without the quark loop intersection with the shock wave \label{quarkloops2a}.}
\end{figure}

\begin{figure}
\includegraphics[width=0.6\textwidth]{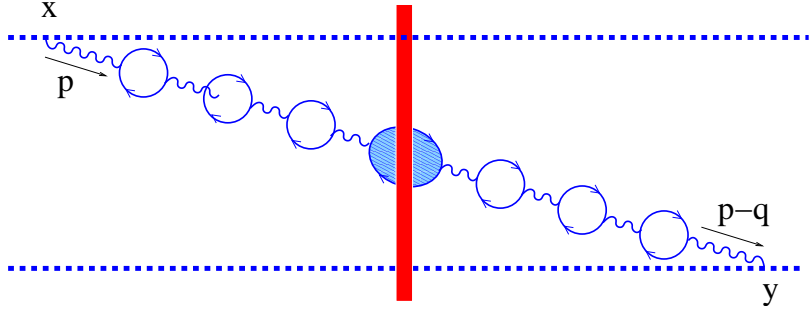}
\caption{Bubble chain with quark loop crossing the shock wave\label{quarkloops2b}.}
\end{figure}

To get an argument of coupling constant we can trace the quark part of the $\beta$-function (proportional to $n_f$). In the leading log approximation the quark part 
 of the $\beta$-function comes from the bubble chain of quark loops in the shock-wave background (cf. Ref \onlinecite{smallxren}). 
 We can either have no intersection of quark loop with the shock wave (see  Fig. \ref{quarkloops2a})
 or we may have one of the loops in the shock-wave background.
 (see Fig. \ref{quarkloops2b})

It is easy to see that the sum of these diagrams yields
\begin{eqnarray}
&&\hspace{-16mm}
{d\over d\eta}{\rm Tr}\{U_xU^\dagger_y\}~=~2\alpha_s{\rm Tr}\{t^aU_xt^bU^\dagger_y\}
\int\! \dhd^2p\dhd^2q~[e^{i(p,x)_\perp}-e^{i(p,y)_\perp}]
[e^{-i(p-q,x)_\perp}-e^{-i(p-q,y)_\perp}]
\nonumber\\
&&\hspace{-16mm}
\times~
{1\over p^2(1+{\alpha_s\over 6\pi}\ln{\mu^2\over p^2})}
\Big(1-{\alpha_sn_f\over 6\pi}\ln{q^2\over \mu^2}\Big)\partial_\perp^2U^{ab}(q)
{1\over (p-q)^2(1+{\alpha_s\over 6\pi}\ln{\mu^2\over (p-q)^2})}
\label{bubblechain1}
\end{eqnarray}
where we have left only the UV part (\ref{uvklad}) of the quark loop.
(In principle, one should also include the dressing of the UV-finite $1/N_c$ term
in Eq. (\ref{nlobklin}) by bubble chain, but I think that it is a separate contribution 
which has nothing to do with the argument of the BK equation).
Replacing the quark part of the $\beta$-function 
$-{\alpha_s\over 6\pi}n_f\ln{p^2\over\mu^2}$ by
the total contribution ${\alpha_s\over 4\pi}b\ln{p^2\over\mu^2}$ (where 
$b-{11\over 3 N_c}-{2\over 3}n_f$), we get
\begin{eqnarray}
&&\hspace{-16mm}
{d\over d\eta}{\rm Tr}\{U_xU^\dagger_y\}~=~2{\rm Tr}\{t^aU_xt^bU^\dagger_y\}
\nonumber\\
&&\hspace{-16mm}
\times\int\! \dhd^2p\dhd^2q~[e^{i(p,x)_\perp}-e^{i(p,y)_\perp}]
[e^{-i(p-q,x)_\perp}-e^{-i(p-q,y)_\perp}]
{\alpha_s(p^2)\over p^2}
\alpha_s^{-1}(q^2)\partial_\perp^2U^{ab}(q)
{\alpha_s((p-q)^2)\over (p-q)^2}
\label{bubblechain2}
\end{eqnarray}

To go to the coordinate space, let us expand the coupling constants in Eq. (\ref{bubblechain2}) 
in powers of $\alpha_s=\alpha_s(\mu^2)$, i.e. return back to Eq. (\ref{bubblechain1})
with ${\alpha_s\over 6\pi}n_f\rightarrow -b{\alpha_s\over 4\pi}$. In the first order we get the UV part of the NLO BK equation (\ref{nlobk})

\begin{eqnarray}
&&\hspace{-6mm}
{d\over d\eta}{\rm Tr}\{U_x U^{\dagger}_y\}~
\label{arg1}\\
&&\hspace{-6mm}
=~{\alpha_s\over 2\pi^2}
\!\int\!d^2z~[{\rm Tr}\{U_x U^{\dagger}_z\}{\rm Tr}\{U_z U^{\dagger}_y\}
-N_c{\rm Tr}\{U_x U^{\dagger}_y\}]
\Big[{(x-y)^2\over X^2 Y^2}
\Big(1+b{\alpha_s \over 4\pi}\ln (x-y)^2\mu^2\Big)-b
{\alpha_s \over 4\pi}{X^2-Y^2\over X^2Y^2}\ln{X^2\over Y^2}\Big]
\nonumber
\end{eqnarray}
In this section, we perform the calculations in the leading log approximation
\begin{equation}
\alpha_s\ln {p^2\over\mu^2}\sim 1, \alpha_s\ll 1
\label{llog}
\end{equation}
 (hence we omit the constant term 
($\sim {5\over 3}$) from the Eq. (\ref{nlobk})).

In the second order in the expansion we obtain
\begin{eqnarray}
&&\hspace{-6mm}
-2\alpha_s\!\int\! \dhd^2p\dhd^2q~e^{i(p,x)_\perp-i(p-q,y)_\perp}
{1\over p^2(1-{b\alpha_s\over 4\pi}\ln{\mu^2\over p^2})}
\Big(1-{b\alpha_s\over 4\pi}\ln{\mu^2\over q^2}\Big)\partial_\perp^2U^{ab}(q)
{1\over (p-q)^2(1-{b\alpha_s\over 4\pi}\ln{\mu^2\over (p-q)^2})}~=
\nonumber\\
&&\hspace{-6mm}
-{b^2\alpha^2_s\over 8\pi^2}(x|
{\ln^2{\mu^2\over p^2}\over p^2}\partial_\perp^2U{1\over p^2}
+{\ln{\mu^2\over p^2}\over p^2}\partial_\perp^2U{\ln{\mu^2\over p^2}\over p^2}
+{1\over p^2}\partial_\perp^2U{\ln^2{\mu^2\over p^2}\over p^2}
-{\ln{\mu^2\over p^2}\over p^2}\Big(\ln{\mu^2\over -\partial_\perp^2}\partial_\perp^2U\Big){1\over p^2}
-{1\over p^2}\Big(\ln{\mu^2\over -\partial_\perp^2}\partial_\perp^2U\Big){\ln{\mu^2\over p^2}\over p^2}
|y)
\nonumber
\end{eqnarray}
Again, it is convenient to replace $\partial_\perp^2U_z$ by $\partial_\perp^2\tilde{U}_z$ 
where $\tilde{U}_z=U_z-{U_x\over 2}-{U_y\over 2}$:

Using the formulas
\begin{eqnarray}
&&\hspace{-26mm}
(x|-\ln^2{\mu^2\over p^2}\tilde{U}{1\over p^2}
-{1\over p^2}\tilde{U}\ln^2{\mu^2\over p^2}
+\ln{\mu^2\over p^2}\Big(\ln{\mu^2\over -\partial_\perp^2}\tilde{U}\Big){1\over p^2}
+{1\over p^2}\Big(\ln{\mu^2\over -\partial_\perp^2}\tilde{U}\Big)\ln{\mu^2\over p^2}
\nonumber\\
&&\hspace{-26mm}
-~\ln{\mu^2\over p^2}\tilde{U}{\ln{\mu^2\over p^2}\over p^2}
-{\ln{\mu^2\over p^2}\over p^2}\tilde{U}\ln{\mu^2\over p^2}|y)
+(x|{\ln{\mu^2\over p^2}\over p^2}|y)\Big[
\Big(\ln{\mu^2\over -\partial_\perp^2}\tilde{U}\Big)_x
+\Big(\ln{\mu^2\over -\partial_\perp^2}\tilde{U}\Big)_y\Big]
\nonumber\\
&&\hspace{-26mm}
=\int\! dz~\Big\{
{1\over 4\pi^2X^2}\Big[\ln X^2\mu^2\ln Y^2\mu^2-\ln{X^2\over Y^2}\ln\Delta^2 \mu^2
-\ln^2\Delta^2 \mu^2\Big]
\nonumber\\
&&\hspace{-26mm}+~
{1\over 4\pi^2Y^2}\Big[\ln X^2\mu^2\ln Y^2\mu^2+\ln{X^2\over Y^2}\ln\Delta^2 \mu^2
-\ln^2\Delta^2 \mu^2\Big]
\Big\}
(U_z-\half U_x-\half U_y)
\nonumber
\end{eqnarray}
and
\begin{eqnarray}
&&\hspace{-6mm}
(x|2{p_i\ln^2{\mu^2\over p^2}\over p^2}\tilde{U}{p_i\over p^2}
+2{p_i\ln{\mu^2\over p^2}\over p^2}\tilde
{U}{p_i\ln{\mu^2\over p^2}\over p^2}+
2{p_i\over p^2}\tilde{U}{p_i\ln^2{\mu^2\over p^2}\over p^2}
-2{p_i\ln{\mu^2\over p^2}\over p^2}\Big(\ln{\mu^2\over 
-\partial_\perp^2}U\Big){p_i\over p^2}
\nonumber\\
&&\hspace{-6mm}
-~2{p_i\over p^2}\Big(\ln{\mu^2\over -\partial_\perp^2}U\Big)
{p_i\ln{\mu^2\over p^2}\over p^2}|y)
=~\int\! {dz\over 4\pi^2}~(U_z-\half U_x-\half U_y)\Bigg\{
\Big[{1\over X^2}+{1\over Y^2}\Big]
\ln^2 \Delta^2\mu^2
\nonumber\\
&&\hspace{-6mm}
+~{\Delta^2\over X^2Y^2}\Bigg[-\ln^2 \Delta^2\mu^2
+{2\Delta_i\over \pi\Delta^2}\int\! dr~\Big[{r_i Y^2-(Y,r)Y_i\over r^2(Y-r)^2}\ln(r-X)^2
-{r_i X^2-(X,r)X_i\over r^2(X-r)^2}\ln(r-Y)^2\Big]   \Bigg]\Bigg\}
\label{flawithintegral}
\end{eqnarray}
we obtain
\begin{eqnarray}
&&\hspace{-6mm}
{d\over d\eta}{\rm Tr}\{U_x U^{\dagger}_y\}^{(2)}~
=~{\alpha_s\over 2\pi^2}\Big({b\alpha_s\over 4\pi}\Big)^2
\!\int\!d^2z~[{\rm Tr}\{U_x U^{\dagger}_z\}{\rm Tr}\{U_z U^{\dagger}_y\}
-N_c{\rm Tr}\{U_x U^{\dagger}_y\}]
\label{arg2}\\
&&\hspace{-6mm}
\times~\Big[{(x-y)^2\over X^2 Y^2}
\ln^2 (x-y)^2\mu^2+{1\over X^2}\ln{X^2\over Y^2}(\ln X^2\mu^2+\ln(x-y)^2\mu^2)
-{1\over Y^2}\ln{X^2\over Y^2}(\ln Y^2\mu^2+\ln(x-y)^2\mu^2)\Big]
\nonumber
\end{eqnarray}
We have omitted the contribution of the last integral in r.h.s. of Eq. 
(\ref{flawithintegral}) since it is negligible in the limits $X\gg Y$, $Y\gg X$ and 
$X,Y\gg x-y$, and therefore can be dropped in the leading log approximation (\ref{llog}). Adding the first-order contribution
(first line in the Eq. (\ref{nlobk}), we get
\begin{eqnarray}
&&\hspace{-6mm}
{d\over d\eta}{\rm Tr}\{U_x U^{\dagger}_y\}~
=~{\alpha_s\over 2\pi^2}
\!\int\!d^2z~[{\rm Tr}\{U_x U^{\dagger}_z\}{\rm Tr}\{U_z U^{\dagger}_y\}
-N_c{\rm Tr}\{U_x U^{\dagger}_y\}]\Big\{{(x-y)^2\over X^2 Y^2}
\Big[1+{b\alpha_s\over 4\pi}\ln (x-y)^2\mu^2
\nonumber\\
&&\hspace{-6mm}
+~\Big({b\alpha_s\over 4\pi}\Big)^2\ln^2 (x-y)^2\mu^2\Big]
+{b\alpha_s\over 4\pi}{1\over X^2}\ln{X^2\over Y^2}\Big[1+{b\alpha_s\over 4\pi}\ln (x-y)^2\mu^2+
{b\alpha_s\over 4\pi}\ln X^2\mu^2\Big]\Big]
\nonumber\\
&&\hspace{-6mm}
-~{b\alpha_s\over 4\pi}{1\over Y^2}\ln{X^2\over Y^2}\Big[1+{b\alpha_s\over 4\pi}\ln (x-y)^2\mu^2+
{b\alpha_s\over 4\pi}\ln Y^2\mu^2\Big]\Big\}
\label{logkva}
\end{eqnarray}

Our guess for the argument of the coupling constant in all orders in $\ln p^2/\mu^2$ is
\footnote{In principle, it should be supported by the (numerical) analysis of third (and higher) higher orders in  $\ln p^2/\mu^2$}
\begin{eqnarray}
&&\hspace{-6mm}
{d\over d\eta}{\rm Tr}\{U_x U^{\dagger}_y\}~
=~{\alpha_s((x-y)^2)\over 2\pi^2}
\!\int\!d^2z~[{\rm Tr}\{U_x U^{\dagger}_z\}{\rm Tr}\{U_z U^{\dagger}_y\}
-N_c{\rm Tr}\{U_x U^{\dagger}_y\}]
\label{arguess}\\
&&\hspace{-6mm}
\times~\Big[{(x-y)^2\over X^2 Y^2}
+{1\over X^2}\Big({\alpha_s(X^2)\over\alpha_s(Y^2)}-1\Big)
+{1\over Y^2}\Big({\alpha_s(Y^2)\over\alpha_s(X^2)}-1\Big)\Big]
\nonumber
\end{eqnarray}
We see now that the argument of the coupling constant in the BK equation
is size of the original dipole $(x-y)^2$ as it was advertised in Eq. (\ref{bkarg}).

\section{Conclusions and Outlook}
 First, there are no new operators at the one-loop level - just as at the tree level, the high-energy scattering can be described in terms 
of Wilson lines. 
The fact that there are no new operators at the one - loop level is rather remarkable. 
In the case of the usual light-cone operator expansion  this is not true - for example,
 if we have the operator
\begin{equation}
[GG]= \int\! du dv ~\theta(u-v)~[\infty p_1,up_1]_xG_{\bullet i}(up_1+x_\perp)[up_1,vp_1]_x
G_\bullet^{~i}(vp_1+x_\perp)[vp_1,-\infty p_1]_x
\label{gg}
\end{equation}
in the leading order, one should expect  the operator 
\begin{equation}
\alpha_s\int\! du dv ~\theta(u-v)\ln(u-v)~[\infty p_1,up_1]_xG_{\bullet i}(up_1+x_\perp)[up_1,vp_1]_x
G_\bullet^{~i}(vp_1+x_\perp)[vp_1,-\infty p_1]_x
\end{equation}
 in the NLO
(in general,  any new loop brings an additional factor $\alpha_s\ln (u-v)$).
 This does not happen here, and in addition the operator $[GG]$ 
appears only in the combination $-i[DG]+[GG]=\partial_\perp^2U$, exactly as at the tree level. I have checked this by the explicit calculation of the quark-loop contribution 
and expect to confirm it by the calculation of the gluon loop. 

 Second conclusion of the paper is that the argument of the coupling constant 
 in the BK equation (obtained from the renormalon-based arguments) appears to be the size  of the parent dipole rather than the size of produced dipoles.
I  have obtained the result for the argument of the coupling constant in the non-linear
evolution of dipoles using the quark part of the $\beta$-function. It is necessary  to confirm this result by calculating the diagrams with gluon loops. 
Also, it would be extremely interesting to check how (and if) this argument of the coupling constant arises from the correlation function of the original dipole and the 
``diamond'' high-energy effective action\cite{diamond} formulated in terms of 
the (renorm-invariant) Wilson lines.
The study is in progress.

\section*{Acknowledgments}
The author is indebted to Yu. Kovchegov for numerous discussions and 
for informing about the results of similar calculation prior to the publication.
The author would like to thank  E. Iancu and other members of theory group at CEA Saclay for for valuable discussions and kind hospitality. 
This work was supported by contract
 DE-AC05-06OR23177 under which the Jefferson Science Associates, LLC operate the Thomas Jefferson National Accelerator Facility.

\section{Appendix: Light-cone expansion of the  quark-loop contribution to gluon propagator in the background field}
\begin{figure}
\includegraphics[width=0.7\textwidth]{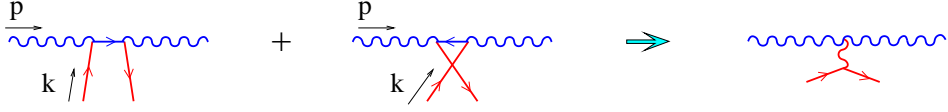}
\caption{``Target'' contribution to the gluon propagator in the external field \label{calo}.}
\end{figure}

The expression (\ref{vesvkladlikone}) should be compared to the light-cone expansion
of the quark-loop part of the gluon propagator in an external field. The quark-loop contribution to the propagator of a gluon in the external field has the form:
\begin{equation}
\hspace{-0mm}
A^m_\bullet(x)A^n_\bullet(y)=\!\int\!\! dx' dy'~(x|{1\over P^2g_{\bullet\mu}+2iG_{\bullet\mu}
+{\cal O}_{\bullet\mu}}|x')^{ma}
{\rm Tr}\{t^a\gamma_\mu
(x'|{1\over\not\! P}|y')t^b\gamma_\nu(y'|{1\over\not\! P}|x')\}
(y'|{1\over P^2g_{\nu\bullet}+2iG_{\nu\bullet}
+{\cal O}_{\nu\bullet}}|y)^{bn}
\label{flaxz1}
\end{equation}
where 
\begin{equation}
{\cal O}_{\mu\bullet}={\cal O}_{\bullet\mu}~=~{1\over\alpha}D^iG_{i\mu}={2p_{2\mu}\over \alpha s}D^iG_{i\bullet}
\label{calomubu}
\end{equation}
An additional term in the gluon propagator is due to the fact that the external gluon
field of the target satisfies the  Yang-Mills equation with a source 
$D^\mu G^a_{\mu\nu}~=~-g\bar{\psi}\gamma_\nu t^a\psi$. 
From the viewpoint of Feynman diagrams in the bF gauge, 
this term comes from the diagrams with the quark insertions shown 
in Fig. \ref{calo} (in the light-like gauge this term arises automatically, see Ref. \onlinecite{npb96}).
For the contribution $\sim {\cal O}_{\mu\bullet}$ the quark propagator reduces to 
\begin{eqnarray}
&&\hspace{-3mm}
\gamma_\mu t^a{(\beta_p+\beta_k)\!\not\! p_2
+(\not\! p+\not\! k)_\perp
\over(\alpha_p+\alpha_k)(\beta_p+\beta_k)s-(p+k)_\perp^2}t^b\!\!\not\! p_1
-t^b\!\!\not\! p_1{(\beta_p-\beta_k)\!\not\! p_2
+(\not\! p-\not\! k)_\perp
\over(\alpha_p-\alpha_k)(\beta_p-\beta_k)s-(p-k)_\perp^2} t^a \gamma_\mu
\label{dens}
\end{eqnarray}
As explained in Ref. \onlinecite{mobzor} at $\alpha_k\ll \alpha_p$ one can 
shift the contour of integration over $\beta_p$ away from the pole in the denominators in 
the above equation. After that  $\beta_p\sim {p_\perp^2\over\alpha_p s}$ so 
one can neglect the terms proportional to transverse momenta in the denominator and in the
numerator. One obtains
\begin{eqnarray}
&&\hspace{-3mm}
{1\over\alpha s} \gamma_\mu\!\not\! p_2\!\not\! p_1t^at^b
-t^bt^a\!\!\not\! p_1\!\not\! p_2 \gamma_\mu~\simeq~{1\over\alpha}if^{abc}t^c \gamma_\mu
\label{calomu}
\end{eqnarray}
which corresponds to the vertex of the insertion of ${\cal O}_{\mu\bullet}$ operator.

We need to expand the Eq. (\ref{flaxz1}) near the 
near the light cone $x\rightarrow y$
and compare it to the light-cone expansion of the same propagator in the shock-wave background (ref{vesvkladlikone}).
The technique for the light-cone expansion of propagators
in external fields was developed in Refs. \onlinecite{pl83,eveq}).
The expansion of the tree-level quark propagator has the form\cite{eveq}:
\begin{eqnarray}
&&\hspace{-6mm}
(x'|{1\over\not\!{P}}|y')
~=~{\not\!\!\Delta\Gamma(2-\epsilon)\over 2\pi^2(-\Delta^2)^{2-\epsilon}}
 +{\Gamma(1-\epsilon)\over  16\pi^2(-\Delta^2)^{1-\epsilon}}
\!\int_0^1\! du~
\Big\{\bar{u}\not\!\!\Delta\sigma G(x_u)+u\sigma G(x_u)
\not\!\!\Delta-2i\bar{u} u\not\!\!\Delta D^\lambda G_{\lambda \Delta}
\nonumber\\
&&\hspace{-6mm}
+~4\not\!\!\Delta\!\int_0^u\! dv~\bar{u} vG_\Delta^{~\xi}(x_u)G_{\Delta \xi}(x_v)\Big\}
\nonumber \\
&&\hspace{-6mm}
+~{\Gamma(-\epsilon)\over  16\pi^2(-\Delta^2)^{-\epsilon}}
\!\int_0^1\! du~\Big\{i(\half -\bar{u} u)D^\lambda G_{\lambda\rho}\gamma^\rho
+{i\over 2}\bar{u} u(1-2u)\hat{D}D^\lambda G_{\lambda\Delta}
-\bar{u} u (1-2u)\gamma^\rho[G_\Delta^{~\xi}(x_u),G_{\rho\xi}(x_u)]
\nonumber\\
&&\hspace{-6mm}
-~\half\bar{u} u\epsilon_{\Delta\mu\nu\lambda}D^\mu D_\xi G^{\xi\nu}\gamma^\lambda\gamma_5
+i{\bar{u} u\over 2}([G_{\Delta\xi}(x_u),\tilde{G}_\nu^{~\xi}(x_u)]+
[\tilde{G}_{\Delta\xi}(x_u),G_\nu^{~\xi}(x_u)])\gamma_\nu\gamma_5
\nonumber \\
&&\hspace{-6mm}
+\int_0^u\! dv~\Big[(\bar{u}-2\bar{u} v-\half)G_{\Delta\xi}(x_u)G^{\lambda\xi}(x_v)\gamma_\lambda
+(v-2\bar{u} v-\half)G^{\lambda\xi}(x_u)G_{\Delta\xi}(x_v)\gamma_\lambda
\nonumber \\
&&\hspace{-6mm}
+~{i\over 2}(1-2\bar{u} -2v)G_{\Delta\xi}(x_u)
\tilde{G}^{\lambda\xi}(x_v)\gamma_\lambda\gamma_5
-{i\over 2}\tilde{G}_{\Delta\xi}(x_u)G^{\lambda\xi}(x_v)\gamma_\lambda\gamma_5\Big]
\Big\}+O({\rm twist ~3})
\label{kvprop}
\end{eqnarray}
where $\Delta=x'-y'$ and  $\epsilon=1-{d_\perp\over 2}$. Hereafter, we use the
notations $x_u\equiv ux'+\bar{u}y'$ and  $G_{\Delta\mu}\equiv G_{\alpha\mu}\Delta^\alpha$ for brevity. 
\footnote{ The terms proportional to $\gamma_5$ may cause problems 
in the dimensional regularization so one needs to return to the previous 
expression for the quark propagator with three antisymmetrized $\gamma$-matrices. 
However, for the particular contributions which we are interested in the product of 
two quark propagators at $d_\perp\neq 2$ is the same as the product of two expressions  (\ref{kvprop}). }

We need to multiply this by similar expansion for the antiquark propagator.  The product of the two quark propagators has the form:
\begin{equation}
{\rm Tr}~t^a\gamma^\alpha(x'|{1\over\!\not\!P}|y')t^b\gamma^\beta(y'|{1\over\!\not\!P}|x')
~=~(\tilde{T}_1)_{\alpha\beta}^{ab}+(\tilde{T}_2)_{\alpha\beta}^{ab}+(\tilde{T}_3)_{\alpha\beta}^{ab}
\label{product}
\end{equation}
where
\begin{eqnarray}
&&\hspace{-6mm}
(\tilde{T}_1)_{\alpha\beta}^{ab}~=~i{B(2-\epsilon,2-\epsilon)\over 4\pi^2}[x',y']^{ab}
(x'|(p^2 g_{\alpha\beta}-p_\alpha p_\beta){\Gamma(\epsilon)\over(-p^2)^{\epsilon}}|y')~
\nonumber\\
&&\hspace{-6mm}
+~{B(2-\epsilon,1-\epsilon)\over 8\pi^4}
{\Gamma(3-2\epsilon)\over  (-\Delta^2)^{3-2\epsilon}}
\!\int_0^1\! du~\Big([x',x_u](2i\bar{u} \Delta_\alpha G_{\beta\Delta}(x_u)
-2iu \Delta_\beta G_{\alpha\Delta}(x_u)+i\Delta^2 G_{\alpha\beta}(x_u))[x_u,y']\Big)^{ab}
\nonumber\\
\label{tildet1}
\end{eqnarray}
\begin{eqnarray}
&&\hspace{-6mm}
(\tilde{T}_2)_{\alpha\beta}^{ab}~=~i{B(2-\epsilon,1-\epsilon)\over 8\pi^4}
{\Gamma(3-2\epsilon)\over  (-\Delta^2)^{3-2\epsilon}}
\!\int_0^1\! du~\bar{u} u(2\Delta_\alpha\Delta_\beta
-\Delta^2 g_{\alpha\beta})\Big([x',x_u]D^\lambda G_{\lambda\Delta})(x_u)[x_u,y']\Big)^{ab}
\nonumber\\
&&\hspace{-6mm}
+~{B(2-\epsilon,-\epsilon)\over 16\pi^4}
{\Gamma(2-2\epsilon)\over (\Delta^2)^{2-2\epsilon}}
\!\int_0^1\! du\Big([x',x_u][\{[-i(\half -\bar{u} u)
D^\lambda G_{\lambda\alpha}(x_u)\Delta_\beta
-~{i\over 2}\bar{u} u(1-2u)D_\alpha D^\lambda G_{\lambda\Delta}(x_u)\Delta_\beta+\alpha\leftrightarrow\beta]
\nonumber \\
&&\hspace{-6mm}
+~i(\half -\bar{u} u)g_{\alpha\beta}D^\lambda G_{\lambda\Delta}(x_u)
+~{i\over 2}\bar{u} u(1-2u)g_{\alpha\beta}(\Delta\cdot D) 
D^\lambda G_{\lambda\Delta}(x_u)\}[x_u,y']\Big)^{ab}
\label{tildet2}
\end{eqnarray}
\begin{eqnarray}
&&\hspace{-6mm}
(\tilde{T}_3)_{\alpha\beta}^{ab}~=-~(2\Delta_\alpha\Delta_\beta-\Delta^2 g_{\alpha\beta})
{B(2-\epsilon,1-\epsilon)\over 2\pi^4}
{\Gamma(3-2\epsilon)\over  (-\Delta^2)^{3-2\epsilon}}
\!\int_0^1\! du dv~{\rm Tr}\Big\{\bar{u} v\theta(u-v)t^b[y',x']t^a[x',x_u]
\nonumber \\
&&\hspace{-6mm}
\times~
G_{\Delta\xi}(x_u)[x_u,x_v]G_\Delta^{~\xi}(x_v)[x_v,y']
+u\bar{v}\theta(v-u)t^a[x',y']t^b[y',x_u]G_{\Delta\xi}(x_u)[x_u,x_v]G_\Delta^{~\xi}(x_v)[x_v,x']
\Big\}
\nonumber\\
&&\hspace{-6mm}
+~{B(2-\epsilon,-\epsilon)\over 16\pi^4}
{\Gamma(2-\epsilon)\over (-\Delta^2)^{2-\epsilon}}
\!\int_0^1\! du dv~{\rm Tr}\Big\{
-2(\Delta_\alpha \delta^\lambda_\beta+\Delta_\beta \delta^\lambda_\alpha
-g_{\alpha\beta}\Delta^\lambda)\Big(\theta(u-v)t^b[y',x']t^a['x,x_u]
\nonumber\\
&&\hspace{-6mm}
\times~\{(\bar{u} -2\bar{u} v-\half)G_{\Delta\xi}(x_u)[x_u,x_v]G_\lambda^{~\xi}(x_v)
+(v-2\bar{u} v-\half)G_\lambda^{~\xi}(x_u)[x_u,x_v]G_{\Delta\xi}(x_v)\}[x_v,y']
\nonumber\\
&&\hspace{-6mm}
+~\theta(v-u)t^a[x',y']t^b[y',x_u]\{(u -2\bar{v} u-\half)
G_{\Delta\xi}(x_u)[x_u,x_v]G_\lambda^{~\xi}(x_v)
+(\bar{v}-2\bar{v} u-\half)G_\lambda^{~\xi}(x_u)[x_u,x_v]
G_{\Delta\xi}(x_v)\}[x_v,x']\Big\}
\nonumber\\
&&\hspace{-6mm}
-~{B(1-\epsilon,1-\epsilon)\over 16\pi^4}
{\Gamma(2-2\epsilon)\over (\Delta^2)^{2-2\epsilon}}
\!\int_0^1\! du dv~{\rm Tr}
\Big\{\Delta^2t^a([x',x_u]G_{\alpha\xi}(x_u)[x_u,y']t^b[y',x_v]G_\beta^{~\xi}(x_v)[x_v,x']
+\alpha\leftrightarrow\beta)+2(\bar{u} v+\bar{v} u)
\nonumber\\
&&\hspace{-6mm}
\times~t^a\Big(g_{\alpha\beta}[x',x_u]G_{\Delta\xi}(x_u)[x_u,y]t^b[y',x_v]
G_\Delta^{~\xi}(x_v)[x_v,x]
-[x',x_u](G_{\Delta\alpha}(x_u)[x_u,y]t^b[y',x_v]G_{\Delta\beta}x_(v)
+\alpha\leftrightarrow\beta)[x_v,x']\Big)
\nonumber\\
&&\hspace{-6mm}
-~2\bar{u} \Delta_\alpha t^a[x',x_u]G_{\Delta\xi}(x_u)
[x_u,y']t^b[y',x_v]G_\beta^{~\xi}(x_v)[x_v,x']
-2\bar{v} \Delta_\alpha t^a[x',x_u]G_{\beta\xi}(x_u)
[x_u,y']t^b[y',x_v]G_\Delta^{~\xi}(v)[x_v,x']
\nonumber\\
&&\hspace{-6mm}
-~2u \Delta_\beta t^a[x',x_u]G_{\Delta\xi}(x_u)
[x_u,y']t^b[y',x_v]G_\alpha^{~\xi}(x_v)[x_v,x']
-2v \Delta_\beta t^a[x',x_u]G_{\alpha\xi}(x_u)
[x_u,y']t^b[y',x_v]G_\Delta^{~\xi}(x_v)[x_v,x']\Big\}
\label{tildet3}
\end{eqnarray}
where we have omitted terms $\sim\epsilon_{\alpha\beta\mu\nu}\Delta^\mu(...)^\nu$ 
and $[G_{\Delta\xi}, G_{\nu}^{~\xi}]$ which do not 
contribute to Eq. (\ref{flaxz1}) with our accuracy.

Next we need to substitute the product (\ref{product}) into the expression (\ref{flaxz1}). Since we will integrate
the expression (\ref{flaxz1}) over $x_\ast$ and $y_\ast$ (to get $U_xU^\dagger_y$) 
we can neglect the terms proportional to $P_\alpha(...)_\beta$ and $(...)_\alpha P_\beta$. Indeed,
using the identity 
\begin{equation}
P_\alpha{1\over P^2g_{\alpha\beta}+2iG_{\alpha\beta}}={1\over P^2}P_\beta
+{1\over P^2}D^\mu G_{\mu\alpha}{1\over P^2g_{\alpha\beta}+2iG_{\alpha\beta}}
\label{flaxz2}
\end{equation}
we get
\begin{equation}
P_\alpha{1\over P^2g_{\alpha\bullet}+2iG_{\alpha\bullet}
+{\cal O}_{\alpha\bullet}}={1\over P^2}P_\bullet,
~~~~{1\over P^2g_{\bullet\alpha}+2iG_{\bullet\alpha}
+{\cal O}_{\bullet\alpha}}P_\alpha=P_\bullet{1\over P^2}
\label{flaxz2}
\end{equation}
As explained in Ref. \onlinecite{mobzor}, one can drop the
terms
proportional to $P_\bullet$  since they 
lead to the terms
proportional to the integral of total derivative, namely
\begin{eqnarray}
&&\int dx_\ast [\infty p_1,{2\over s}x_\ast p_1]t^{a}
[{2\over s}x_\ast p_1,-\infty p_1](D_\bullet \Phi
({2\over s}x_\ast p_1,...))_{ab}\nonumber\\
&=&\int dx_\ast \frac {d}{dx_\ast}\{[\infty p_1,{2\over s}x_\ast p_1]t^{a}
[{2\over s}x_\ast p_1,-\infty p_1]
(\Phi ({2\over s}x_\ast p_1,...))_{ab}\}=0
\label{izobzora}
\end{eqnarray}
Using this property one can rewrite Eq. (\ref{product}) in the form
\begin{equation}
{\rm Tr}~t^a\gamma^\alpha(x'|{1\over\!\not\!P}|y')t^b\gamma^\beta(y'|{1\over\!\not\!P}|x')
~\Leftrightarrow~(T_1)_{\alpha\beta}^{ab}+(T_2)_{\alpha\beta}^{ab}+(T_3)_{\alpha\beta}^{ab}
\label{product}
\end{equation}
where
\begin{eqnarray}
\hspace{-3mm}
(T_1)_{\alpha\beta}^{ab}
&=&ig_{\alpha\beta}{\Gamma^2(2-\epsilon)\over 4\pi^2\Gamma(4-2\epsilon)}
(x|P^2{\Gamma(\epsilon)\over(-P^2)^{\epsilon}}|y)^{ab}-{g\Gamma^2(2-\epsilon)\over 4\pi^4\Gamma(3-2\epsilon)}(x| {\Gamma(\epsilon)\over(-p^2)^{\epsilon}}|y)\!\!
\int_0^1\!\! du~([x',x_u]G_{\alpha\beta}(x_u)[x_u,y'])^{ab}
\nonumber\\
&+&{ig\over 8\pi^2}{\Gamma(3-\epsilon)\Gamma(1-\epsilon)\over\Gamma(4-2\epsilon)}
(x| {\Gamma(\epsilon)p_\rho\over(-p^2)^{\epsilon}}|y)\!\!
\int_0^1\!\! du~([x',x_u][\bar{u} \delta^\rho_\alpha G_{\Delta\beta}(x_u)-
u\delta^\rho_\beta G_{\Delta\alpha}(x_u)][x_u,y'])^{ab}
\nonumber\\
(T_2)_{\alpha\beta}^{ab}
&=&
g_{\alpha\beta}(1-2\epsilon)
{g\Gamma^2(2-\epsilon)\over 4\pi^2\Gamma(4-2\epsilon)}(x|{\Gamma(1+\epsilon)\over(-p^2)^{1+\epsilon}}|y)
\!\int_0^1\!\!  du~([x',x_u]D^\mu G^{ab}_{\mu\Delta}(x_u)[x_u,y'])^{ab}
+(\tilde{T}_2)_{\alpha\beta}^{ab}
\label{Ts}\\
(T_3)_{\alpha\beta}^{ab}
&=&
ig_{\alpha\beta}(1-2\epsilon){g^2\Gamma^2(2-\epsilon)\over 2\pi^2\Gamma(4-2\epsilon)}
(x|{\Gamma(1+\epsilon)\over(-p^2)^{1+\epsilon}}|y)
\!\int_0^1\!\!  du\!
\!\int_0^u\!\! \!dv~\bar{u} v([x',x_u]G_{\Delta\xi}(x_u)[x_u,x_v]G_\Delta^{~\xi}(x_v)[x_v,y'])^{ab}
+(\tilde{T}_3)_{\alpha\beta}^{ab}
\nonumber
\end{eqnarray}
and $\Leftrightarrow$ means ``equal up to the contributions  $\sim~P_\alpha(...)_\beta$ and $(...)_\alpha P_\beta$''.

\begin{figure}
\includegraphics[width=0.99\textwidth]{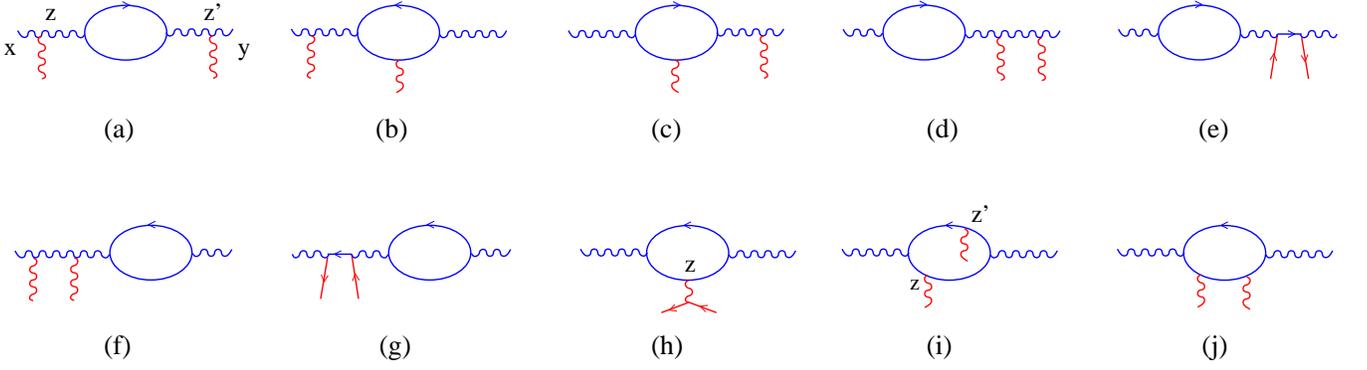}
\caption{Quark-loop contribution to the gluon propagator in an external field \label{kvnlo14}.}
\end{figure}

Next we expand the propagator (\ref{flaxz1}) near the light cone. The first contribution comes from the $T_1$ term which represents diagrams in Fig. \ref{kvnlo14} a-g.
The calculation yields
\begin{eqnarray}
&&\hspace{-0mm}
\!\int\!\! dx' dy'~(x|{1\over P^2g_{\bullet\mu}+2iG_{\bullet\mu}
+{\cal O}_{\bullet\mu}}|x')^{ma}
T_{1\alpha\beta}^{ab}
(y'|{1\over P^2g_{\nu\bullet}+2iG_{\nu\bullet}
+{\cal O}_{\nu\bullet}}|y)^{bn}
\nonumber\\
&&\hspace{-6mm}
=~{1\over 8\pi^2}B(2-\epsilon,2-\epsilon)(x|{1\over p^2}
\Big\{{\partial_\perp^2U^{ab}\over\alpha},{\Gamma(\epsilon)\over(-p^2)^\epsilon}\Big\}
{1\over p^2}|y)
-{ig\over 4\pi^2}B(2-\epsilon,2-\epsilon)(x|{1\over p^2}
\Big\{{D^iG_{i\bu}^{ab}\over\alpha},{\Gamma(\epsilon)\over(-p^2)^\epsilon}\Big\}
{1\over p^2}|y)
\nonumber\\
&&\hspace{-6mm}
+~{g^2\Gamma^2(1-\epsilon)\over 32\pi^2\Gamma(4-2\epsilon)}\!\int_0^\infty\! {\dhd\alpha\over \alpha^3}\!\int_{y_\ast}^{x_\ast}\!\!
d{2\over s}z_\ast \!\int_{y_\ast}^{z_\ast}\!\! d{2\over s}z'_\ast~
([x_\ast,z_\ast]_xG_{\bu i}(z_\ast,x_\perp)[z_\ast,z'_\ast]_x
G_{\bu j}(z'_\ast,x_\perp)[z'_\ast,y_\ast]_x)^{ab}
\nonumber\\
&&\hspace{-6mm}\times~
\Big({4i\over\alpha s}\Big)^\epsilon
\Bigg\{g_{ij}\Big[
 {1\over\epsilon}[(2-\epsilon)(x-z')_\ast^\epsilon-(x-z)_\ast^\epsilon-(z-z')_\ast^\epsilon]
+  {1\over\epsilon}[(2-\epsilon)(z-y)_\ast^\epsilon-(z'-y)_\ast^\epsilon-
(z-z')_\ast^\epsilon]
   \nonumber\\
&&\hspace{-6mm}
+~ 2(z-z')_\ast^\epsilon-
2(2-\epsilon){(z-z')_\ast\over(x-z')_\ast^{1-\epsilon}}
+(1-\epsilon) {(z-z')_\ast^2\over(x-z')_\ast^{2-\epsilon}}
 -2(2-\epsilon){(z-z')_\ast\over (z-y)_\ast^{1-\epsilon}}+
  (1-\epsilon) {(z-z')_\ast^2\over(z-i)_\ast^{2-\epsilon}}
\nonumber\\
&&\hspace{-6mm}
+~{2(1-\epsilon)\over3-\epsilon}(z-z')_\ast^2[\Delta_\ast^{-2+\epsilon}+(z-z')_\ast^{-2+\epsilon}
-(x-z)_\ast^{-2+\epsilon}-(z'-y)_\ast^{-2+\epsilon}]\Big]
\nonumber\\
&&\hspace{-6mm}
+~{1\over (1+\epsilon)}
\Big[(x-z)_\ast^{1+\epsilon}+(z-z')_\ast^{1+\epsilon}
-(x-z')_\ast^{1+\epsilon}
+(1+\epsilon){(x-z)_\ast(z-z')_\ast\over(x-z')_\ast^{1-\epsilon}}
+(z'-y)_\ast^{1+\epsilon}+(z-z')_\ast^{1+\epsilon}
\nonumber\\
&&\hspace{-6mm}-~
(z-y)_\ast^{1+\epsilon}+(1+\epsilon){(z-z')_\ast(z'-i)_\ast\over(x-z')_\ast^{1-\epsilon}}\Big]
\Big({(3-\epsilon)g_{ij}\over \Delta_\ast}+i{\alpha s\over 4\Delta_\ast^2}
[\Delta_\perp^2g_{ij}-\Delta_i\Delta_j]\Big)
\nonumber\\
&&\hspace{-6mm}+~{g_{ij}\over\epsilon} (1-\epsilon)(3-2\epsilon)
 [2(z-z')_\ast^\epsilon+(x-z)_\ast^\epsilon+(z-y)_\ast^\epsilon-2(x-z')_\ast^\epsilon-2(z'-y)_\ast^\epsilon]
 \Bigg\}~\Big(-{i\alpha s\over 4\pi\Delta_\ast}\Big)^{1-\epsilon}
e^{i{\Delta_\perp^2\over 4\Delta_\ast}\alpha s}
\label{vkladot1}
\end{eqnarray}
In our ``external'' field the characteristic distances $z_\ast(z'_\ast)$ are of the order of width of the shock wave: $z_\ast,z'_\ast\sim e^{\eta_2}\sqrt{s/m^2}$. As we shall see below, the characteristic distances $x_\ast$ and $y_\ast$ are 
$\sim e^{\eta_2}\sqrt{s/m^2}$
so we can neglect $z_\ast$ and $z'_\ast$ in comparison to  $x_\ast$ and/or $y_\ast$.
The formula (\ref{vkladot1}) simplifies to
\begin{eqnarray}
&&\hspace{-0mm}
\!\int\!\! dx' dy'~(x|{1\over P^2g_{\bullet\mu}+2iG_{\bullet\mu}
+{\cal O}_{\bullet\mu}}|x')^{ma}
T_{1\alpha\beta}^{ab}
(y'|{1\over P^2g_{\nu\bullet}+2iG_{\nu\bullet}
+{\cal O}_{\nu\bullet}}|y)^{bn}
\nonumber\\
&&\hspace{-6mm}
=~{1\over 8\pi^2}B(2-\epsilon,2-\epsilon)(x|{1\over p^2}
\Big\{{\partial_\perp^2U^{ab}\over\alpha},{\Gamma(\epsilon)\over(-p^2)^\epsilon}\Big\}
{1\over p^2}|y)
+{ig^3\over 16\pi^2}{B(2-\epsilon,2-\epsilon)\over\epsilon}\!\int_0^\infty\! {\dhd\alpha\over \alpha^3}\nonumber\\
&&\hspace{-6mm}\times~
\Big({4i\over\alpha s}\Big)^\epsilon
[x_\ast^\epsilon+ (-y)_\ast^\epsilon]~
 \Big(-{i\alpha s\over 4\pi\Delta_\ast}\Big)^{1-\epsilon}
e^{i{\Delta_\perp^2\over 4\Delta_\ast}\alpha s}\!\int_{y_\ast}^{x_\ast}\!\!
d{2\over s}z_\ast ~
([x_\ast,z_\ast]_xD^iG_{i\bu}(z_\ast,x_\perp)[z_\ast,z'_\ast]_x
[z_\ast,y_\ast]_x)^{ab}
\nonumber\\
&&\hspace{-6mm}
+~{g^2\Gamma(1-\epsilon)\Gamma(2-\epsilon)\over 16\pi^2\Gamma(4-2\epsilon)}\!\int_0^\infty\! {\dhd\alpha\over \alpha^3}\!\int_{y_\ast}^{x_\ast}\!\!
d{2\over s}z_\ast \!\int_{y_\ast}^{z_\ast}\!\! d{2\over s}z'_\ast~
([x_\ast,z_\ast]_xG_\bu^{~i}(z_\ast,x_\perp)[z_\ast,z'_\ast]_x
G_{\bu i}(z'_\ast,x_\perp)[z'_\ast,y_\ast]_x)^{ab}
\nonumber\\
&&\hspace{-6mm}\times~
\Big({4i\over\alpha s}\Big)^\epsilon
\Big\{
- {1-\epsilon\over\epsilon}[x_\ast^\epsilon
+ (-y)_\ast^\epsilon-2(z-z')_\ast^\epsilon]
+{1\over 3-\epsilon}(z-z')_\ast^{\epsilon}
 \Big\}~
 \Big(-{i\alpha s\over 4\pi\Delta_\ast}\Big)^{1-\epsilon}
e^{i{\Delta_\perp^2\over 4\Delta_\ast}\alpha s}
\label{vkladot1simpl}
\end{eqnarray}

The term $\sim T_2$ coming from the diagram in Fig. \ref{kvnlo14}h has the form
\begin{eqnarray}
&&\hspace{-0mm}
\!\int\!\! dx' dy'~(x|{1\over p^2}|x')
T_{2\bullet\bullet}^{ab}
(y'|{1\over p^2}|y)
\label{vkladot2}\\
&&\hspace{-6mm}
=~-{i\over 32\pi^2}
\!\int_0^\infty\!{\dhd\alpha\over\alpha^3}\Big({i\over\alpha s}\Big)^{\epsilon}
 \Big(-{i\alpha s\over 4\pi\Delta_\ast}\Big)^{1-\epsilon}
e^{i{\Delta_\perp^2\over 4\Delta_\ast}\alpha s}
~{\Gamma(2-\epsilon)\Gamma(1-\epsilon)\over\Gamma(3-2\epsilon)}
\Bigg\{
{(4-\epsilon)(1-\epsilon)\over \epsilon(2-\epsilon)(3-\epsilon)}
[x_\ast^{\epsilon}+(-y)_\ast^{\epsilon}-\Delta_\ast^{\epsilon}-\epsilon x_\ast y_\ast\Delta_\ast^{\epsilon -2}]
\nonumber\\
&&\hspace{-6mm}
-~{2(2-2\epsilon+\epsilon^2)\over\epsilon(2-\epsilon)(1-\epsilon^2)}
\Big[{1-\epsilon\over \Delta_\ast}+{i\alpha s\Delta_\perp^2\over 4\Delta_\ast^2}\Big]  
[x_\ast^{1+\epsilon}+(-y)_\ast^{1+\epsilon}-\Delta_\ast^{1+\epsilon}-{(1+\epsilon)x_\ast y_\ast\over\Delta_\ast^{1-\epsilon}}]
-{1\over \epsilon(1-\epsilon)(2+\epsilon)}
\Big[{(2-\epsilon)(1-\epsilon)\over \Delta_\ast^2}
\nonumber\\
&&\hspace{-6mm}
+~{i(2-\epsilon)\alpha s\over 2\Delta_\ast^3}\Delta_\perp^2
-{\alpha^2s^2\over 16\Delta_\ast^4}\Delta_\perp^4\Big]
[x_\ast^{2+\epsilon}+(-y)_\ast^{2+\epsilon}
-\Delta_\ast^{2+\epsilon}-(2+\epsilon)x_\ast y_\ast\Delta_\ast^{\epsilon}]
\Bigg\}
\!\int_{y_\ast}^{x_\ast}\!\!
d{2\over s}z_\ast ([x_\ast,z_\ast]_xD^\lambda G_{\lambda\bu}(z_\ast,x_\perp)
[z_\ast,y_\ast]_x^{ab}
\nonumber
\end{eqnarray}
where we have neglected $z_\ast$ in comparison to $x_\ast,y_\ast$ as discussed above.
Since there is no field outside the shock wave,
\footnote{Up to a possible pure gauge field which does not change the result of the analysis, see the discussion in \cite{mobzor})}
at $x_\ast y_\ast>0$ the contribution (\ref{vkladot2}) vanishes and at $x_\ast>0,y_\ast<0$ one can extend the limits of integration over $z_\ast$ to 
$\pm\infty$ and get 
\begin{eqnarray}
&&\hspace{-0mm}
\!\int\!\! dx' dy'~(x|{1\over p^2}|x')
T_{2\bullet\bullet}^{ab}
(y'|{1\over p^2}|y)
\label{vkladot2red}\\
&&\hspace{-6mm}
\stackrel{x_\ast>0>y_\ast}{=}~-{i\over 32\pi^2}
\!\int_0^\infty\!{\dhd\alpha\over\alpha^3}\Big({i\over\alpha s}\Big)^{\epsilon}
 \Big(-{i\alpha s\over 4\pi\Delta_\ast}\Big)^{1-\epsilon}
e^{i{\Delta_\perp^2\over 4\Delta_\ast}\alpha s}
~{\Gamma(2-\epsilon)\Gamma(1-\epsilon)\over\epsilon\Gamma(3-2\epsilon)}
\Big\{
{(4-\epsilon)(1-\epsilon)\over (2-\epsilon)(3-\epsilon)}
[x_\ast^{\epsilon}+(-y)_\ast^{\epsilon}
\nonumber\\
&&\hspace{-6mm}
-~\Delta_\ast^{\epsilon}-\epsilon x_\ast y_\ast\Delta_\ast^{\epsilon -2}]
-{2(2-2\epsilon+\epsilon^2)\over (2-\epsilon)(1-\epsilon^2)}
\Big[{1-\epsilon\over \Delta_\ast}+{i\alpha s\Delta_\perp^2\over 4\Delta_\ast^2}\Big]  
[x_\ast^{1+\epsilon}+(-y)_\ast^{1+\epsilon}-\Delta_\ast^{1+\epsilon}-{(1+\epsilon)x_\ast y_\ast\over\Delta_\ast^{1-\epsilon}}]
\nonumber\\
&&\hspace{-6mm}-~
{1\over (1-\epsilon)(2+\epsilon)}
\Big[{(2-\epsilon)(1-\epsilon)\over \Delta_\ast^2}
+{i(2-\epsilon)\alpha s\over 2\Delta_\ast^3}\Delta_\perp^2
-{\alpha^2s^2\over 16\Delta_\ast^4}\Delta_\perp^4\Big]
[x_\ast^{2+\epsilon}+(-y)_\ast^{2+\epsilon}
-\Delta_\ast^{2+\epsilon}-(2+\epsilon)x_\ast y_\ast\Delta_\ast^{\epsilon}]
\Big\}~[DG]_x^{ab}
\nonumber
\end{eqnarray}
where $[DG]$ is defined by Eq. (\ref{dgigg}). Similarly, the integral 
$\int_{y_\ast}^{x_\ast}\!\!
d{2\over s}z_\ast ([x_\ast,z_\ast]_xD^\lambda G_{\lambda\bu}(z_\ast,x_\perp)$ 
in the second term in the r.h.s. of Eq. (\ref{vkladot1simpl}) can be reduced to $[DG]_x$.
The case $x_\ast<0,y_\ast>0$ is obtained from (\ref{vkladot2red}) by the substitution $x\leftrightarrow y$.

The contribution of diagrams in Fig. \ref{kvnlo14} i,j is
\begin{eqnarray}
&&\hspace{-1mm}
\!\int\!\! dx' dy'~(x|{1\over p^2}|x')
T_{3\bullet\bullet}^{ab}
(y'|{1\over p^2}|y)
\label{vkladot3}\\
&&\hspace{-1mm}
=~{g^2\over 16\pi^2}B(1-\epsilon,2-\epsilon)
\!\int_0^\infty\!{\dhd\alpha\over\alpha^3}\Big({4i\over\alpha s}\Big)^{\epsilon}
 \Big(-{i\alpha s\over 4\pi\Delta_\ast}\Big)^{1-\epsilon}
e^{i{\Delta_\perp^2\over 4\Delta_\ast}\alpha s}
\!\int_{y_\ast}^{x_\ast}\!\!
d{2\over s}z_\ast \!\int_{y_\ast}^{z_\ast}\!\! d{2\over s}z'_\ast
\nonumber\\
&&\hspace{-1mm}\times~
\Bigg(\Bigg\{
{(4-\epsilon)(1-\epsilon)\over\epsilon(2-\epsilon)(3-\epsilon)}[x_\ast^{\epsilon}+(-y)_\ast^{\epsilon}
-\Delta_\ast^{\epsilon}-(z-z')_\ast^{\epsilon}]
-{(z-z')_\ast^{\epsilon}\over (2-\epsilon)(3-\epsilon)}-{(4-\epsilon)(1-\epsilon)x_\ast y_\ast \over (2-\epsilon)(3-\epsilon)\Delta_\ast^{2-\epsilon}}
\nonumber\\
&&\hspace{-1mm}
-~\Big[{1-\epsilon\over \Delta_\ast}+{i\alpha s\Delta_\perp^2\over 4\Delta_\ast^2}\Big]   {2(2-2\epsilon+\epsilon^2)\over \epsilon(1-\epsilon^2)(2-\epsilon)}
[x_\ast^{1+\epsilon}+(-y)_\ast^{1+\epsilon}
-~\Delta_\ast^{1+\epsilon}-(1+\epsilon)x_\ast y_\ast\Delta_\ast^{\epsilon-1}]
\nonumber\\
&&\hspace{-1mm}
-~\Big[{(2-\epsilon)(1-\epsilon)\over \Delta_\ast^2}
+{i(2-\epsilon)\alpha s\over 2\Delta_\ast^3}\Delta_\perp^2
-{\alpha^2s^2\over 16\Delta_\ast^4}\Delta_\perp^4\Big]
{[x_\ast^{2+\epsilon}+(-y)_\ast^{2+\epsilon}
-\Delta_\ast^{2+\epsilon}-(2+\epsilon)x_\ast y_\ast\Delta_\ast^{\epsilon}]\over
\epsilon(1-\epsilon)(2+\epsilon)}
\Bigg\}
\nonumber\\
&&\hspace{26mm}\times~([x_\ast,z_\ast]_xG_{\bu i}(z_\ast,x_\perp)
[z_\ast,z'_\ast]_xG_\bu^{~i}(z'_\ast,x_\perp)[z'_\ast,y_\ast])^{ab}
\nonumber\\
&&\hspace{-1mm}
+~\Bigg\{-g_{ij}{x_\ast y_\ast \over \Delta_\ast^{2-\epsilon}}
-~2\Big[{1-\epsilon\over \Delta_\ast}+{i\alpha s\Delta_\perp^2\over 4\Delta_\ast^2}\Big]  g_{ij}{2-2\epsilon+\epsilon^2\over \epsilon(1-\epsilon^2)(2-\epsilon)}
[x_\ast^{1+\epsilon}+(-y)_\ast^{1+\epsilon}-\Delta_\ast^{1+\epsilon}-(1+\epsilon)x_\ast y_\ast\Delta_\ast^{\epsilon-1}]
\nonumber\\
&&\hspace{-1mm}
-~\Big[{(2-\epsilon)(1-\epsilon)\over \Delta_\ast^2}
+{i(2-\epsilon)\alpha s\over 2\Delta_\ast^3}\Delta_\perp^2
-{\alpha^2s^2\over 16\Delta_\ast^4}\Delta_\perp^4\Big]
{[x_\ast^{2+\epsilon}+(-y)_\ast^{2+\epsilon}
-\Delta_\ast^{2+\epsilon}-(2+\epsilon)x_\ast y_\ast\Delta_\ast^{\epsilon}]
\over \epsilon(1-\epsilon)(2+\epsilon)}g_{ij}
\nonumber\\
&&\hspace{-1mm}
-~\Big[{1-\epsilon\over \Delta_\ast}+{i\alpha s\Delta_\perp^2\over 4\Delta_\ast^2}\Big]  g_{ij}
{x_\ast^{1+\epsilon}+(-y)_\ast^{1+\epsilon}-\Delta_\ast^{1+\epsilon}\over \epsilon(1+\epsilon)}
-\Big[2{g_{ij}\over \Delta_\ast}
-{i\alpha s\over \Delta_\ast^2}\Delta_i\Delta_j\Big]
{x_\ast^{1+\epsilon}+(-y_\ast)^{1+\epsilon}-\Delta_\ast^{1+\epsilon}-(1+\epsilon)
x_\ast y_\ast \Delta_\ast^{\epsilon-1}\over (1-\epsilon^2)(2-\epsilon)}
\Bigg\}
\nonumber\\
&&\hspace{16mm}
\times~[x_\ast,z_\ast]^{am}2{\rm Tr}\{t^mG_{\bu i}(z_\ast,x_\perp)
[z_\ast,z'_\ast]t^nG_{\bu j}(z'_\ast,x_\perp)[z'_\ast,z_\ast]_x
+m\leftrightarrow n\}[z_\ast,y_\ast]_x^{nb}\Bigg)
\nonumber
\end{eqnarray}
The final expression for the light-cone expansion of the quark-loop contribution to 
gluon propagator in the sum of the expressions (\ref{vkladot1simpl}), 
(\ref{vkladot2}), and (\ref{vkladot3}). A very important observation is that 
the contributions proportional to
\begin{equation}
g^4n_f \int\!dz_\ast\!\int dz'_\ast~\theta(z-z')~(z-z')^\epsilon G_{\bu i}(z_\ast)G_{\bu i}(z'_\ast)
\label{badterm}
\end{equation}
present in the Eqs. (\ref{vkladot1simpl}) and (\ref{vkladot3}) cancel in their sum. If it were not true, there would be an addtional contribution to the gluon propagator (\ref{gluprop}) at the $g^4$ level coming from the small-size (large-momenta) 
quark loop. Indeed, the calculations of Feynman diagrams with the propagators (\ref{gluprop}) and (\ref{kvpropagator}) implies that we first take limit $z_\ast,z'_\ast\rightarrow 0$ and limit $d_\perp\rightarrow 2$ afterwards. 
With such order of limits, the contribution (\ref{badterm}) vanishes. However, 
the proper order of these limits is to take at first $d_\perp\rightarrow 2$ (which 
will give finite expressions after adding the counterterms) and then try to impose the
condition that the external field is very narrow by taking the limit $z_\ast, z'_\ast\rightarrow 0$. In this case, Eq. (\ref{badterm}) reduces to 
$g^4n_f[GG]$.
 The non-commutativity of these limits would mean that 
the contribution ${1\over p^2}[GG]{1\over p^2}$ should be added to the gluon propagator (\ref{gluprop}) to restore the correct result. Fortunately, the terms $\sim$
(\ref{badterm}) cancel which means that there are no additional contributions to the 
gluon propagator coming from the quark loop inside the shock wave ($\equiv$ quark loop
with large momenta).
 
Since there is no external field outside the shock wave,  after cancellation of the terms $\sim (z-z')^\epsilon$ we see that at $x_\ast y_\ast>0$ the sum 
of Eq. (\ref{vkladot1simpl}) and Eq. (\ref{vkladot3}) vanishes, and at  $x_\ast>0>y_\ast$ one can extend the limits of integration in the gluon operators
to $\pm\infty$ and obtain 
\begin{eqnarray}
&&\hspace{-3mm}
 \int_{y_\ast}^{x_\ast}\!\!\!dz_\ast\!\int_{y_\ast}^{x_\ast}\!\!\!dz'_\ast~ 
  ([x_\ast,z_\ast]_x G_{\bu i}(z_\ast,x_\perp)[z_\ast,z'_\ast]_xG_{\bu i}(z'_\ast,x_\perp)
 [z'_\ast,y_\ast]_x)^{ab}~\rightarrow~[GG]^{ab}_x
\\
&&\hspace{-3mm}
 \int_{y_\ast}^{x_\ast}\!\!\!dz_\ast\!\int_{y_\ast}^{z_\ast}\!\!\!dz'_\ast~
 [x_\ast,z_\ast]^{am}{\rm Tr}\{t^mG_{\bu i}(z_\ast,x_\perp)
[z_\ast,z'_\ast]t^nG_{\bu j}(z'_\ast,x_\perp)[z'_\ast,z_\ast]_x
+m\leftrightarrow n\}[z_\ast,y_\ast]_x^{nb}
~\rightarrow~{\rm Tr}\{t^a\partial_iU_xt^b\partial_jU^\dagger_x\}
\nonumber
\end{eqnarray}

We get
\begin{eqnarray}
&&\hspace{-1mm}
\!\int\!\! dx' dy'~(x|{1\over P^2g_{\bullet\alpha}+2iG_{\bullet\alpha}
+{\cal O}_{\bullet\alpha}}|x')^{ma}
(T_{1\alpha\beta}^{ab}+T_{2\alpha\beta}^{ab}+T_{3\alpha\beta}^{ab})
(y'|{1\over P^2g_{\beta\bullet}+2iG_{\beta\bullet}
+{\cal O}_{\beta\bullet}}|y)^{bn}
\nonumber\\
&&\hspace{-1mm}
\stackrel{x_\ast>0>y_\ast}{=}~{1\over 8\pi^2}B(2-\epsilon,2-\epsilon)(x|{1\over p^2}
\Big\{{\partial_\perp^2U^{ab}\over\alpha},{\Gamma(\epsilon)\over(-p^2)^\epsilon}\Big\}
{1\over p^2}|y)
~+~{g^2\over 32\pi^2}B(1-\epsilon,2-\epsilon)
\!\int_0^\infty\!{\dhd\alpha\over\alpha^3}\Big({4i\over\alpha s}\Big)^{\epsilon}
\nonumber\\
&&\hspace{-1mm}\times~
 \Big(-{i\alpha s\over 4\pi\Delta_\ast}\Big)^{1-\epsilon}
e^{i{\Delta_\perp^2\over 4\Delta_\ast}\alpha s}\Bigg(\Bigg\{- {1-\epsilon\over\epsilon(3-2\epsilon)}[x_\ast^\epsilon
+ (-y)_\ast^\epsilon]+
{(4-\epsilon)(1-\epsilon)\over\epsilon(2-\epsilon)(3-\epsilon)}[x_\ast^{\epsilon}+(-y)_\ast^{\epsilon}
-\Delta_\ast^{\epsilon}-\epsilon x_\ast y_\ast\Delta_\ast^{\epsilon -2}]
\nonumber\\
&&\hspace{-1mm}
-~\Big[{1-\epsilon\over \Delta_\ast}+{i\alpha s\Delta_\perp^2\over 4\Delta_\ast^2}\Big]   {2(2-2\epsilon+\epsilon^2)\over \epsilon(1-\epsilon^2)(2-\epsilon)}
[x_\ast^{1+\epsilon}+(-y)_\ast^{1+\epsilon}
-~\Delta_\ast^{1+\epsilon}-(1+\epsilon)x_\ast y_\ast\Delta_\ast^{\epsilon-1}]
\nonumber\\
&&\hspace{-1mm}
-~\Big[{(2-\epsilon)(1-\epsilon)\over \Delta_\ast^2}
+{i(2-\epsilon)\alpha s\over 2\Delta_\ast^3}\Delta_\perp^2
-{\alpha^2s^2\over 16\Delta_\ast^4}\Delta_\perp^4\Big]
{[x_\ast^{2+\epsilon}+(-y)_\ast^{2+\epsilon}
-\Delta_\ast^{2+\epsilon}-(2+\epsilon)x_\ast y_\ast\Delta_\ast^{\epsilon}]\over
\epsilon(1-\epsilon)(2+\epsilon)}\Bigg\}~\partial_\perp^2U_x^{ab}
\nonumber\\
&&\hspace{-1mm}
+~\Bigg\{-{x_\ast y_\ast \over \Delta_\ast^{2-\epsilon}}
-~2\Big[{1-\epsilon\over \Delta_\ast}+{i\alpha s\Delta_\perp^2\over 4\Delta_\ast^2}\Big]  {2-2\epsilon+\epsilon^2\over \epsilon(1-\epsilon^2)(2-\epsilon)}
[x_\ast^{1+\epsilon}+(-y)_\ast^{1+\epsilon}-\Delta_\ast^{1+\epsilon}-(1+\epsilon)x_\ast y_\ast\Delta_\ast^{\epsilon-1}]
\nonumber\\
&&\hspace{-1mm}
-~\Big[{(2-\epsilon)(1-\epsilon)\over \Delta_\ast^2}
+{i(2-\epsilon)\alpha s\over 2\Delta_\ast^3}\Delta_\perp^2
-{\alpha^2s^2\over 16\Delta_\ast^4}\Delta_\perp^4\Big]
{[x_\ast^{2+\epsilon}+(-y)_\ast^{2+\epsilon}
-\Delta_\ast^{2+\epsilon}-(2+\epsilon)x_\ast y_\ast\Delta_\ast^{\epsilon}]
\over \epsilon(1-\epsilon)(2+\epsilon)}-\Big[{1-\epsilon\over \Delta_\ast}
\nonumber\\
&&\hspace{-1mm}
+~{i\alpha s\Delta_\perp^2\over 4\Delta_\ast^2}\Big]  
{x_\ast^{1+\epsilon}+(-y)_\ast^{1+\epsilon}-\Delta_\ast^{1+\epsilon}\over \epsilon(1+\epsilon)}
-{2\over \Delta_\ast}
{x_\ast^{1+\epsilon}+(-y_\ast)^{1+\epsilon}-\Delta_\ast^{1+\epsilon}-(1+\epsilon)
x_\ast y_\ast \Delta_\ast^{\epsilon-1}\over (1-\epsilon^2)(2-\epsilon)}
\Bigg\}~4{\rm Tr}\{t^a\partial_iU_xt^b\partial^iU^\dagger_x\}
\nonumber\\
&&\hspace{16mm}+~
{i\alpha s\over \Delta_\ast^2}\Delta^i\Delta^j
{x_\ast^{1+\epsilon}+(-y_\ast)^{1+\epsilon}-\Delta_\ast^{1+\epsilon}-(1+\epsilon)
x_\ast y_\ast \Delta_\ast^{\epsilon-1}\over (1-\epsilon^2)(2-\epsilon)}
~4{\rm Tr}\{t^a\partial_iU_xt^b\partial_jU^\dagger_x\}
\Bigg)
\label{symmats}
\end{eqnarray}
We see that the light-cone expansion  of gluon propagator contains only Wilson lines
and their derivatives as should be expected after cancellation of the ``contaminating''
terms (\ref{badterm}).

Next, to get the expansion of $[\infty,0]_x\otimes[0,-\infty]_y$ near the light cone we integrate the expression (\ref{symmats}) over $x_\ast$  from $0$ to $\infty$ and over $y_\ast$ from $-\infty$ to 0. 
 It is easy to demonstrate that
\begin{eqnarray}
&&\hspace{-0mm}
\int_0^\infty\!\! dx_\ast \!\int_{-\infty}^0\!\!dy_\ast~ \Big(-{i\alpha s\over 4\pi\Delta_\ast}\Big)^{1-\epsilon}
e^{i{\Delta_\perp^2\over 4\Delta_\ast}\alpha s}
~\Bigg\{
{(4-\epsilon)(1-\epsilon)\over \epsilon(2-\epsilon)(3-\epsilon)}
[x_\ast^{\epsilon}+(-y)_\ast^{\epsilon}-\Delta_\ast^{\epsilon}-\epsilon x_\ast y_\ast\Delta_\ast^{\epsilon -2}]
\nonumber\\
&&\hspace{-6mm}
-~{2(2-2\epsilon+\epsilon^2)\over\epsilon(2-\epsilon)(1-\epsilon^2)}
\Big[{1-\epsilon\over \Delta_\ast}+{i\alpha s\Delta_\perp^2\over 4\Delta_\ast^2}\Big]  
[x_\ast^{1+\epsilon}+(-y)_\ast^{1+\epsilon}-\Delta_\ast^{1+\epsilon}-{(1+\epsilon)x_\ast y_\ast\over\Delta_\ast^{1-\epsilon}}]
-{1\over \epsilon(1-\epsilon)(2+\epsilon)}
\Big[{(2-\epsilon)(1-\epsilon)\over \Delta_\ast^2}
\nonumber\\
&&\hspace{-6mm}
+~{i(2-\epsilon)\alpha s\over 2\Delta_\ast^3}\Delta_\perp^2
-{\alpha^2s^2\over 16\Delta_\ast^4}\Delta_\perp^4\Big]
[x_\ast^{2+\epsilon}+(-y)_\ast^{2+\epsilon}
-\Delta_\ast^{2+\epsilon}-(2+\epsilon)x_\ast y_\ast\Delta_\ast^{\epsilon}]
\Bigg\}~=~0
\nonumber
\end{eqnarray}
(in particular, it means that the term (\ref{vkladot2red}) coming from the diagram in Fig. \ref{kvnlo14}h does not contribute). The result of the integration of Eq. (\ref{symmats}) has the form
\begin{eqnarray}
&&\hspace{-1mm}
[\infty,0]_x\otimes[0,-\infty]_y\stackrel{x_\perp\rightarrow y_\perp}{\rightarrow}
~
\nonumber\\
&&\hspace{-1mm}
=~{2\alpha_s^2\over \pi}\Delta\eta~B(2-\epsilon,2-\epsilon)
(x_\perp|{1\over p_\perp^2}
\Big\{\partial_\perp^2U^{ab},{\Gamma(\epsilon)\over(p_\perp^2)^\epsilon}
\Big\}{1\over p_\perp^2}|y_\perp)~
+~{\alpha_s^2\over 8\pi^2}\Delta\eta ~B(1-\epsilon,1-\epsilon)
{\Gamma(-1-2\epsilon)\over (\Delta_\perp^2)^{-1-2\epsilon}}
\label{xz2}\\
&&\hspace{-1mm}
\times~\Big(\Big\{
-{(4-\epsilon)(1-\epsilon)\over 6\epsilon(1+\epsilon)}
+{7+\epsilon\over 6(1+\epsilon)}
-{1\over 3(1+\epsilon)(2+\epsilon)}
\Big\}~{\rm Tr}\{t^a\partial_iU_xt^b\partial_iU^\dagger_x\}
-
{2(1+2\epsilon)\over 3(1+\epsilon)(2+\epsilon)}{\Delta_i\Delta_j\over\Delta_\perp^2}
~{\rm Tr}\{t^a\partial_iU_xt^b\partial_jU^\dagger_x\}
\Big)
\nonumber
\end{eqnarray}
and therefore
\begin{eqnarray}
&&\hspace{-36mm}
\partial^x_i[\infty,0]_x\otimes\partial^y_i[0,-\infty]_y
\stackrel{x_\perp\rightarrow y_\perp}{\rightarrow}
~{\alpha_s^2\Delta\eta\over \pi^2}
{B(2-\epsilon)\over\epsilon}
{\Gamma(-2\epsilon)\over(\Delta_\perp^2)^{-2\epsilon}}
~\partial_\perp^2U^{ab}_x
+~{\alpha_s^2\Delta\eta\over 4\pi^2}~B(1-\epsilon)
{\Gamma(-2\epsilon)\over (\Delta_\perp^2)^{-2\epsilon}}
\nonumber\\
&&\hspace{-36mm}
\times~\Big(\Big\{
-{2\over 3\epsilon}+2-{1\over 3(1+\epsilon)}
\Big\}~2{\rm Tr}\{t^a\partial_iU_xt^b\partial_iU^\dagger_x\}
-{4\epsilon\over 3(1+\epsilon)}{\Delta_i\Delta_j\over\Delta_\perp^2}~2{\rm Tr}\{t^a\partial_iU_xt^b\partial_jU^\dagger_x\}
\Big)
\label{schitaem4}
\end{eqnarray}
so we obtain
\begin{eqnarray}
&&\hspace{-3mm}
{\rm Tr}\{\partial^x_iU_x\partial^y_iU^\dagger_y\}
\stackrel{x_\perp\rightarrow y_\perp}{\rightarrow}
~{\alpha_s^2n_f\over \pi^2}\Delta\eta
{B(2-\epsilon,2-\epsilon)\over\epsilon}
{\Gamma(-2\epsilon)\over(\Delta_\perp^2)^{-2\epsilon}}
~U^{ab}_x\partial_\perp^2U^{ab}_x
+~{\alpha_s^2n_f\over 4\pi^2N_c}\Delta\eta~B(1-\epsilon,1-\epsilon)
{\Gamma(-2\epsilon)\over (\Delta_\perp^2)^{-2\epsilon}}
\nonumber\\
&&\hspace{-3mm}
\times~\Bigg[\Big\{
{2\over 3\epsilon}-2+{1\over 3(1+\epsilon)}
\Big\}\delta_{ij}
+{4\epsilon\over 3(1+\epsilon)}{\Delta_i\Delta_j\over\Delta_\perp^2}
\Bigg]~{\rm Tr}\{\partial_iU_x\partial_jU^\dagger_x\}
\label{patamy}
\end{eqnarray}

Last, we need to write down the sum of $1/\epsilon$ counterterms
to diagrams in Fig. (\ref{kvnlo14}) a-g. It can be read from the first term
in the r.h.s. of Eq. (\ref{vkladot1simpl}): 
\begin{eqnarray}
&&\hspace{-6mm}
-{\alpha_s^2n_f\over 12\pi^2\epsilon}\Delta\eta~{\Gamma(-2\epsilon)\over(\Delta_\perp^2)^{-2\epsilon}}~
U^{ab}_x\partial_\perp^2U^{ab}_x
\label{kterm}
\end{eqnarray}
Adding the counterterm (\ref{kterm}) to Eq. (\ref{patamy}) we get
\begin{eqnarray}
&&\hspace{-3mm}
{\rm Tr}\{\partial^x_iU_x\partial^y_iU^\dagger_y\}
\stackrel{x_\perp\rightarrow y_\perp}{\rightarrow}
~{\alpha_s^2n_f\over \pi^2}\Delta\eta
\Big[{B(2-\epsilon,2-\epsilon)\over\epsilon}
{\Gamma(-2\epsilon)\over(\Delta_\perp^2)^{-2\epsilon}}
-{1\over 12\epsilon}{\Gamma(-2\epsilon)\over(\Delta_\perp^2)^{-2\epsilon}}\Big]
~U^{ab}_x\partial_\perp^2U^{ab}_x
\nonumber\\
&&\hspace{-3mm}
+~{\alpha_s^2n_f\over 4\pi^2N_c}\Delta\eta~B(1-\epsilon,1-\epsilon)
{\Gamma(-2\epsilon)\over (\Delta_\perp^2)^{-2\epsilon}}
\Big[\Big(
{2\over 3\epsilon}-2+{1\over 3(1+\epsilon)}
\Big)\delta_{ij}
+{4\epsilon\over 3(1+\epsilon)}{\Delta_i\Delta_j\over\Delta_\perp^2}
\Big]~{\rm Tr}\{\partial_iU_x\partial_jU^\dagger_x\}
\label{patamy}
\end{eqnarray}
which coincides with Eq. (\ref{vesvkladlikone}).

\end{document}